# Real-World Adoption, Use, and Governance of an LLM-Based Electronic Health Record Search and Synthesis Platform

Jiunn-Tyng Yeh, Michael Gao, Marshall Nichols, Anika Mohan, Tareq Aljurf, Michael Revoir, Matthew Gardner, Bradley Hintze, William Ratliff, Suresh Balu, and Will Knechtle

Duke Institute for Health Innovation, Duke University Health System, Durham, North Carolina

## Abstract

We evaluated the longitudinal adoption and use of Scout, an electronic health record-integrated large language model platform that provides cited search and synthesis of patient records, during staged deployment across a multi-hospital academic health system. Over 11 months, 2,635 users completed 80,465 tasks. 1,768 clinical users initiated use within the first 29 days of system-wide launch, which included a required training. Daily initiation increased 21.2-fold at launch, from 2.9 to 61.0 users per day. Of the 495 users who initiated use during the launch week, 310 did so in the first two days. Use spanned more than 20 recorded specialty categories, five care-practitioner roles, and 46,485 distinct prompts involving retrieval, synthesis, extraction, evaluation, and drafting, with 752 tasks saved for reuse. Among care practitioners who completed the required course, 77.9% went on to use Scout, with a median of 1.7 days after course completion. The highest-intensity user quintile accounted for 59.0% of tasks and 61.9% of tokens. Exploratory prompt analysis showed use across retrieval, synthesis, extraction, evaluation, and note-drafting. These findings suggest that rapid, broad adoption requires clinician-led task design and evaluation, traceable governance proportional to reach, centralized capacity planning and user support, and training integrated with access.

## Introduction

Large language model (LLM) tools are increasingly integrated into electronic health records (EHRs) to support clinical documentation, chart search, longitudinal synthesis, and open-ended queries into patient records.[1-3] Most reports describe institutional deployment, adoption, task performance, and user experience,[4-6] but rarely link longitudinal adoption patterns with repeat use of prompt-defined purposes and tasks, variation across professional roles, and computational demand.

Transitioning from pilot implementation to institution-wide deployment also requires planning for infrastructure capacity, training, incident reporting, workflow validation, and ongoing monitoring.[2,4] Utilization logs can provide evidence on how adoption and demand evolve, what types of tasks users request, and which tasks become reusable workflows.[5,6]

Scout is an LLM-based EHR search and synthesis platform, developed and operated by the Duke Institute for Health Innovation (DIHI). Scout answers a care practitioner's natural-language question about a single patient's record with inline citations linking every claim to its source data. It does not diagnose or recommend treatment, and, as deployed during this study, it did not retain memory across tasks. Scout integrated and advanced two earlier DIHI efforts, an LLM-based chart-interface prototype and a novel pipeline incorporating real-time EHR data, into a citation-first architecture that uses agentic, multi-step retrieval across clinical notes, laboratory results, medications, pathology notes and imaging notes.[7] In a prospective, randomized, evaluator-blinded, two-period crossover trial conducted across seven specialties from September-December 2025, Scout reduced task-completion time by 37.6% (ratio of geometric means, 0.624; 95% CI, 0.565-0.689; P<0.001). It also met prespecified noninferiority criteria for accuracy, completeness, and relevance and demonstrated statistically superior completeness; a manual review of 187 Scout claims found no fabricated content.[7] Following the trial and a limited launch, on July 9, 2026, Scout was made available broadly with integrated access within Epic across a multi-hospital academic health system. Users interact with Scout by entering a task: a natural-language instruction specifying what to do with a patient's record and what output to produce. Tasks may take the form of questions or directives, such as "summarize this admission." The task's text is its prompt; a task saved to the Scout library for reuse is called a SmartTask. Each task is also a logged event tied to one user, one patient record, one response, and one set of citations, thereby framing a Scout session as a bounded operation on a single chart rather than a conversational chatbot. ChatEHR, a comparable EHR-grounded LLM assistant developed at Stanford Medicine, is the closest published analogue and provides an external comparison for adoption speed and use patterns.[6] To describe this limited launch (the non-trial access period before official launch), we characterized adoption and follow-on use, resource concentration, multidimensional prompt-defined use, variation by professional role, and governance during scale-up. Re-evaluation of task performance and assessment of clinical outcomes were outside the scope of this deployment study.

# Methods

## Cohort and setting

We conducted a retrospective observational cohort study of Scout task logs collected from September 10, 2025, through August 6, 2026. Eligible events required a valid user identifier and timestamp. We excluded demonstration sessions, duplicate records, and records with suspected source-schema misalignment, and prompt analyses required nonempty text. During the study period Scout used a single-patient, single-task model. Each task stood alone within one patient's record, with no conversational threading between tasks and no batch execution across patient cohorts, so changes of intent within a thread did not need to be analyzed.

Scout operates within Maestro Care, Duke Health's Epic environment, and can be accessed from a patient chart through the More Activities menu or via Sidebar Activity. Users can search across encounters or within a specific time range, expand data sources to include discrete data outside of clinical notes, and incorporate Care Everywhere data. The user can also auto-populate the prompt with previously saved SmartTasks. Scout returns a written response in which each claim includes citations, and selecting a citation opens an evidence card displaying the source note or data point which supports the claim.

Following a randomized trial, a limited launch, and institutional approvals, 8,306 provider accounts became eligible to access Scout through Maestro Care beginning on July 9, 2026. Access required completion of a tracked learning-management-system course that lasted approximately 11 minutes on average. Eligible providers included physicians and residents; advanced practice providers (APPs), including physician assistants, nurse practitioners, certified registered nurse anesthetists, and certified nurse midwives; and dental, podiatric, and chiropractic care practitioners. Registered nurses were not included in the July 2026 system-wide eligibility expansion, but had used Scout through earlier trial and limited-launch pathways. The longitudinal cohort therefore includes registered nurses and other staff provisioned before system-wide access.

## Measures and analysis

We summarized cumulative and weekly active users, daily first-time users, and monthly return after first use. Power users were defined as the highest quintile by tasks per active day. We compared the token-per-task distribution of power users with that of other users using a two-sided Mann-Whitney U test, and calculated each group's shares of tasks, tokens, and estimated cost. For complete calendar weeks, we summarized median response time, token volume, estimated cost based on Azure OpenAI API prices for models used by Scout. Reported financial metrics reflect late-2025/early-2026 baseline pricing; token costs for frontier models used have since decreased by >70% following Scout's initial deployment.

Each distinct prompt was classified once and its labels were applied to all occurrences. Independent dimensions captured intended use (why), requested information operation (what Scout was asked to do), and output form. Prompt-use mode distinguished SmartTasks, non-SmartTask prompts repeated across patient records, and other prompts; design features could co-occur. A SmartTask is a prompt that a Scout user has saved to the Scout library and recalled by typing a period followed by the task's shortcut name. A physician investigator led development and validation of this exploratory framework using 500 randomly sampled prompts and a post-development review sample; frozen codebooks were applied with GPT-5.6 Luna. Role analyses used each user's recorded role. Governance milestones were abstracted from contemporaneous records. The Supplementary Appendix provides complete methods and additional results, de-identified prompt examples (Tables S1 and S2), and the complete frozen classifier prompts.

This study was conducted under Duke University Health System Institutional Review Board protocol Pro00118622. Participants in the randomized trial and the limited launch provided written informed consent using IRB-approved consent forms specific to each phase. All Scout users, including those who accessed the system during system-wide deployment, acknowledged an institutional use agreement stating that usage would be logged for safety monitoring, quality improvement, and product development, and that users can be contacted for related surveys or studies.

## Setting and staged deployment

Scout was designed, developed, and is operated by the Duke Institute for Health Innovation (DIHI), combining two earlier internal efforts, an LLM chart-interface prototype and a structured and unstructured Epic data-extraction pipeline, into a single citation-first platform with agentic, multistep retrieval across chart data types. From January through April 2026, Scout underwent a formal application security review by Duke's Health Information Security Office, the standard prerequisite to broad deployment of an application handling protected health information, before broader deployment. The review examined data movement and misuse resistance, addressing whether protected information could leave Duke or cross between patients and whether the model could be elicited to produce overtly harmful content. Clinical-content correctness was assessed separately through task-level evaluations. As designed, testers were unable to extract data across patients despite deliberate attempts, and Microsoft Defender for AI correctly detected and alerted on jailbreak attempts. The review also identified findings that DIHI resolved before broader rollout: the standalone, pre-Maestro Care version of Scout relied on Maestro Care as the authoritative source of patient-level authorization rather than independently enforcing record-level access; a nonstandard character-substitution technique could extract more of Scout's system instructions than intended; a single access group exceeding 500 users required a more scalable periodic-review process; and audit logging needed to be made independently verifiable, which DIHI addressed by routing prompt and response logs into Duke's Splunk protected-health-information index. Security approval was granted April 3, 2026 and was soon followed by confirmation from the Duke Compliance office and the Algorithm-Based Clinical Decision Support Oversight Committee on April 27.

Scout progressed to system-wide access through a randomized trial and a limited launch, the latter being the whole period of non-trial access that preceded official launch (Table 1). The randomized trial ran from September 2025 through December 2025. The limited launch began on September 22, 2025, overlapped the trial for its first three months, and continued until system-wide access on July 9, 2026. Throughout this period, access was granted under the research protocol and required consent, training, and collection of user role and interest information. The first phase enrolled 213 experienced DIHI collaborators spanning more than 20 specialties and service lines to provide early user-experience feedback. This group included 82 faculty physicians, 48 residents and postgraduate trainees, 15 advanced practice providers, 23 nurses, and 45 other clinical staff including pharmacists and physical therapists. The advanced-practice-provider category included physician assistants and nurse practitioners, while the nursing category excludes nurse practitioners. This cohort submitted 6,641 tasks and increased weekly active use to approximately 70 users by the end of 2025.[7]

Formal security review and governance approval occupied the first four months of 2026 while limited launch access continued to expand. Through 2026 the limited launch entered a second phase, in which specialty groups were increasingly given access so that they too could provide feedback, and from there access spread by word of mouth: a referral-review and research-screening extension at the Duke Cancer Center added 22 users, and extensions of enrollment added 40 psychiatry and behavioral health users, 16 transplant center users, 14 cardiology users, 6 nephrology users, and 12 medical genetics and genomics users, and a hospital medicine discharge summary effort began in February 2026. Growth from April through early July was primarily driven by word of mouth without formal promotion, and all users continued to join through the research protocol.

**Table 1. Staged deployment of Scout, September 2025 through August 2026.**

| Stage | Dates | Who had access | What it produced |
|---|---|---|---|
| Randomized trial | Sept-Dec 2025 | Enrolled trial participants across 7 specialties | Evidence of shorter real-world task-completion time and noninferior accuracy, completeness, and relevance[7] |
| Limited launch, phase 1: DIHI collaborators | Sept 22-Dec 31, 2025 | Care practitioners and administrators with prior engagement with DIHI innovation projects | Engaged 213 users across 20+ specialties to execute 6,641 tasks, capture early UX insights, and achieve ~70 weekly active users |
| Security and governance review | Jan-Apr 2026 | Limited launch access continued to expand to the cancer center, transplant center, and hospital medicine | Security approval on Apr 3; Compliance and ABCDS confirmation on Apr 27 |
| Limited launch, phase 2: specialty extension | Jan-Jul 8, 2026 | Specialty groups added for feedback (psychiatry, cardiology, nephrology, medical genetics), then word of mouth | Secured 110 additional users under the research protocol framework |
| System-wide Epic access | Jul 9, 2026 onward | 8,306 eligible provider accounts, after required training | User adoption surged to 1,768 in 29 days, averaging 61 daily, a 21.2x growth acceleration. |

*The limited launch denotes the entire non-trial access period preceding official launch, September 22, 2025 through July 8, 2026; it overlapped the randomized trial for its first three months and continued through the security review. The stages were defined retrospectively.*

# Results

## Cohort, adoption, and return use

Of 82,171 source records, 80,465 eligible tasks from 2,635 users were included, representing 37,002 patient records and 46,485 distinct prompts. In total, 1,706 records were excluded. The pipeline identified 1,403 repeated event records and 869 records with suspected source-schema misalignment, and exclusion reasons were not mutually exclusive. Cohort, adoption, and retention measures are summarized in Table 2.

Before system-wide Maestro Care access, 867 users initiated use over 302 days (2.9 per day). During the subsequent 29 days, 1,768 users initiated use (61.0 per day; 0.73% of the launch-eligible population per day), a 21.2-fold unadjusted increase. Uptake during the rollout week was concentrated in the first two days: 310 of 495 new users (62.6%) initiated use on July 9 or 10. Weekly active users increased from 317 before rollout to 826 during rollout and 1,299 the following week despite required training (Figure 1). First-time users increased from 62 to 495 to 620 across the same three weeks. An independently sourced, user-level session export corroborated the shape of this adoption curve and is reported below with the provisioning-to-first-use analysis.

Pooled monthly retention was 59.8% at month 1, 47.5% at month 3, and 45.9% at month 6, fewer cohorts were observable at later time points. Month 1 retention increased from 54.3% for the March cohort to 75.0% for the June cohort. The analysis included 2,475 users with a complete first-use month and excluded 160 users whose first use occurred during partial August. The first-use cohort grew from 92 users in June to 1,722 in July. Cohort-specific month 1 retention ranged from 40.3% to 81.1%. Later estimates were based on progressively fewer cohorts and do not represent comparisons within a fixed population. Because August was incomplete, the July system-wide access cohort did not contribute to month 1 retention.

**Table 2. Cohort, adoption, and return use.**

| Measure | Value |
|---|---|
| Source records screened | 82,171 |
| Eligible task events | 80,465 from 2,635 users |
| Records excluded | 1,706 (1,403 repeated events; 869 suspected schema misalignment; reasons not mutually exclusive) |
| Patient records represented | 37,002 |
| Distinct prompts | 46,485 |
| Pre-launch initiation (Sept 10, 2025-Jul 8, 2026; 302 days) | 867 users; mean 2.9 per day |
| Post-launch initiation (Jul 9-Aug 6, 2026; 29 days) | 1,768 users; mean 61.0 per day; 0.73% of 8,306 eligible accounts per day |
| Unadjusted ratio of daily initiation | 21.2-fold |
| First two days of access (Jul 9-10) | 310 of 495 rollout-week initiations (62.6%) |
| Weekly active users (weeks of Jun 29 / Jul 6 / Jul 13) | 317 / 826 / 1,299 |
| First-time users (weeks of Jun 29 / Jul 6 / Jul 13) | 62 / 495 / 620 |
| Retention analysis population | 2,475 users with a complete first-use month (160 partial-August users excluded) |
| Pooled retention, month 1 | 59.8% (450 of 753) |
| Pooled retention, month 2 | 54.3% (359 of 661) |
| Pooled retention, month 3 | 47.5% (271 of 571) |
| Pooled retention, month 6 | 45.9% (157 of 342) |
| Cohort-specific month-1 retention | Range 40.3%-81.1% (March cohort 54.3%; June cohort 75.0%) |

*Source: Scout task logs.*

**Figure 1. Adoption and monthly cohort retention.** Panel A shows cumulative distinct Scout users, computed directly from session logs highlighting the system-wide Epic access on July 9, 2026. This session-level export totaled 2,534 unique users by August 6, 2026, relative to the 2,635-user task-eligible cohort analyzed across this study. The minor discrepancy of 101 users occurred during the pre-launch period (771 versus 867 prior to July 9, 2026; 1,763 versus 1,768 thereafter). These two datasets were extracted at different implementation stages and for different operational purposes (see Discussion); The task-log cohort serves as the analytical reference standard throughout this manuscript, whereas Panel A uses the session data export to preserve temporal (daily) resolution. Panel B depicts period-specific monthly retention by calendar month of first-use, reconstructed from reported pooled retention percentages because per-cohort longitudinal task data were not available for this figure. Each cell indicates the percentage of the entry cohort with >= 1 eligible task during the corresponding post-index month (Month 0 is 100%). Empty cells signify insufficient follow-up duration; activity in intervening months was not required.

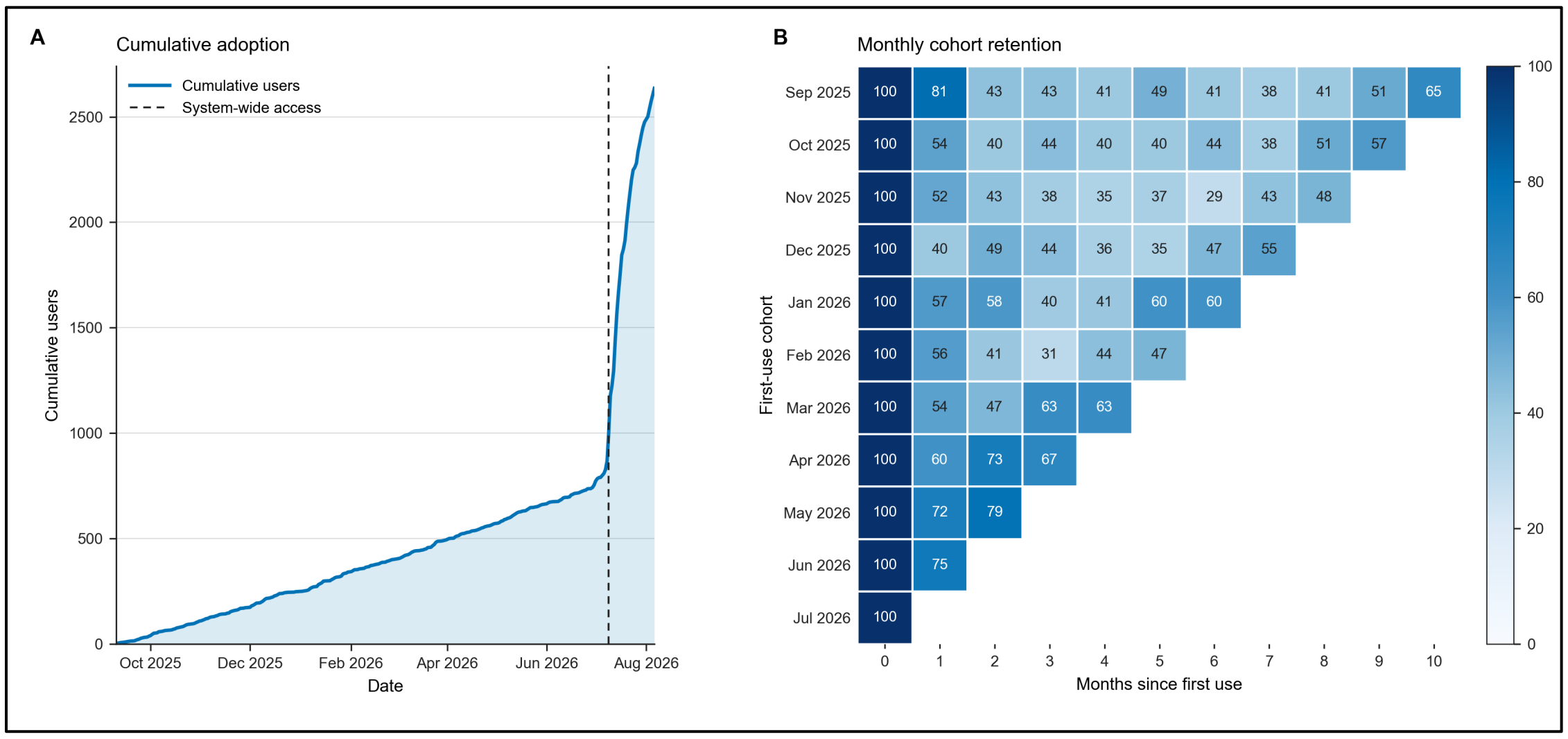

## Use intensity and computational demand

The 527 power users (20.0%) had, by definition, greater intensity: a median 5.0 tasks per active day (IQR, 4.25-6.90) versus 2.0 (IQR, 1.33-2.60) among other users. Median tokens per observed task were 0.63 million versus 0.39 million (P<0.001). Power users were defined by a minimum intensity of 3.77 tasks per active day, and the token-per-task comparison included 520 power users and 1,978 other users with recorded token values. Power users accounted for 59.0% of tasks, 61.9% of observed tokens, and 60.5% of estimated cost (Figure 2; Table 3). Intensity did not imply persistence: power users had a median of 7 active days versus 3 among other users, but 64 power users (12.1%) were active on only 1 day.

**Table 3. Use intensity, power users versus other users.**

| Measure | Power users (n=527) | Other users (n=2,108) |
|---|---|---|
| Tasks per active day, median (IQR) | 5.0 (4.25-6.90) | 2.0 (1.33-2.60) |
| Tokens per observed task, median (IQR), millions | 0.63 (0.27-1.22) | 0.39 (0.11-0.93) |
| Active days, median | 7 | 3 |

| Measure | Power users (n=527) | Other users (n=2,108) |
|---|---|---|
| Users active on only 1 day | 64 (12.1%) | Not reported |
| Eligible tasks | 47,482 (59.0% of 80,465) | 32,983 (41.0%) |
| Observed tokens | 55.7 billion (61.9% of 90.0 billion) | 34.3 billion (38.1%) |
| Estimated model API cost | $79,165 (60.5% of $130,898) | $51,733 (39.5%) |

*Token values were available for 94.4% of events and cost values for 95.4%. Other-user figures are arithmetic complements of the reported power-user totals.*

Across 46 complete weeks, 73,969 tasks generated 84.5 billion observed tokens and an estimated $124,151 in model API cost. Weekly token volume increased before launch, predominantly among returning users, and reached 5.91 billion during rollout, with an estimated cost of $6,664. Median weekly response time remained between 33.3 and 38.1 seconds across the four complete rollout and early-access weeks (Figure 3; Table 4). Demand continued to increase after the rollout week: median weekly volume across those four weeks was 9.18 billion tokens at a median weekly cost of $10,413, peaking in the week beginning July 27 at 10.70 billion tokens and $11,926. Across the full observation period the median response time was 39.9 seconds and the 90th percentile was 153.3 seconds.

Estimated cost per task declined substantially during the observation period. The median cost per task decreased from $1.35 in February 2026 to $0.27 in the first week of August, while the corresponding means decreased from $3.48 to $1.12; over the same period, monthly task volume increased approximately tenfold from February to July 2026. Selected record windows contributed to these differences. Tasks using an unbounded all-time window had a median cost of $1.51 and a mean cost of $3.70, compared with $0.13 and $0.49, respectively, for a 30-day window. The all-time option was phased out between May and July 2026 and replaced with a two-year window; it accounted for 31.7% of tasks in May, 16.4% in June, and none thereafter. Cost per task continued to decline after the observation period, reaching a median of $0.09 and a mean of $0.25 during the 30 days ending September 23, 2026.

The prelaunch increase was driven primarily by existing users. Between the weeks beginning June 15 and June 22, task volume increased by 54.9% while active users increased by 8.8% and first-time users declined from 22 to 19. Returning users contributed 97.7% of token volume during the week beginning June 22 and 93.2% during the week beginning June 29. Cost per million tokens was nearly unchanged across those weeks ($1.135 and $1.137).

**Table 4. Computational demand and response time across 46 complete weeks.**

| Measure | Value |
|---|---|
| Task events analyzed | 73,969 across 46 complete weeks |
| Observed tokens | 84.5 billion (values available for 94.1% of events) |
| Estimated model API cost | $124,151 (values available for 95.3% of events) |
| Response time, median | 39.9 seconds (66,286 events with a recorded response) |
| Response time, 90th percentile | 153.3 seconds |
| Week before rollout (Jun 29) | 3.54 billion tokens; $4,025; median response 36.9 s |

| Measure | Value |
|---|---|
| Rollout week (Jul 6) | 5.91 billion tokens (+67.1%); $6,664 (+65.6%); median response 33.3 s |
| Four rollout and post-access weeks, median | 9.18 billion tokens; $10,413; median response 33.3-38.1 s |
| Peak week (Jul 27) | 10.70 billion tokens; $11,926 |
| Cost per million tokens (weeks of Jun 22 / Jun 29) | $1.135 / $1.137 |

*Source: Scout task logs. Only complete Monday-through-Sunday weeks are included; totals use available observations without imputation.*

**Figure 2. User-level utilization intensity and concentration.** Power users represent the highest quintile by eligible tasks per active day. Panel A represents medians and interquartile ranges for task volume per active day and token consumption per task. Panel B illustrates the relative contributions of power-user vs. remaining users, eligible tasks, tokens, and estimated cost. This metric captures daily utilization rather than longitudinal persistence.

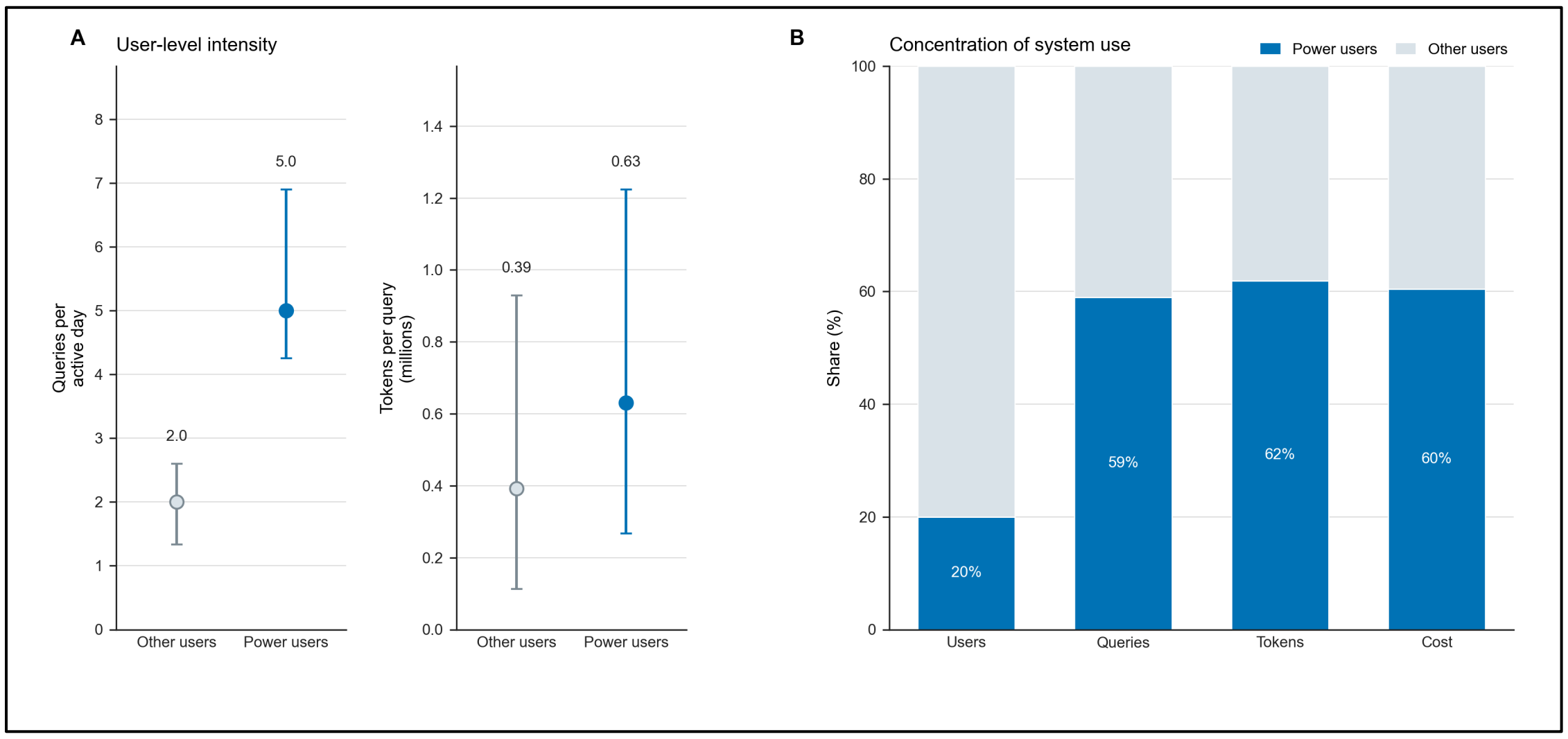

**Figure 3. Weekly computational demand and response time.** Panel A shows weekly token volume, Panel B depicts estimated cost, and Panel C presents median response time for events with confirmed system outputs. The dashed vertical line marks system-wide deployment in Epic on July 9, 2026. Analyses were restricted to complete Monday-through-Sunday calendar weeks, and token volume and cost aggregates were computed from observed data without imputation.

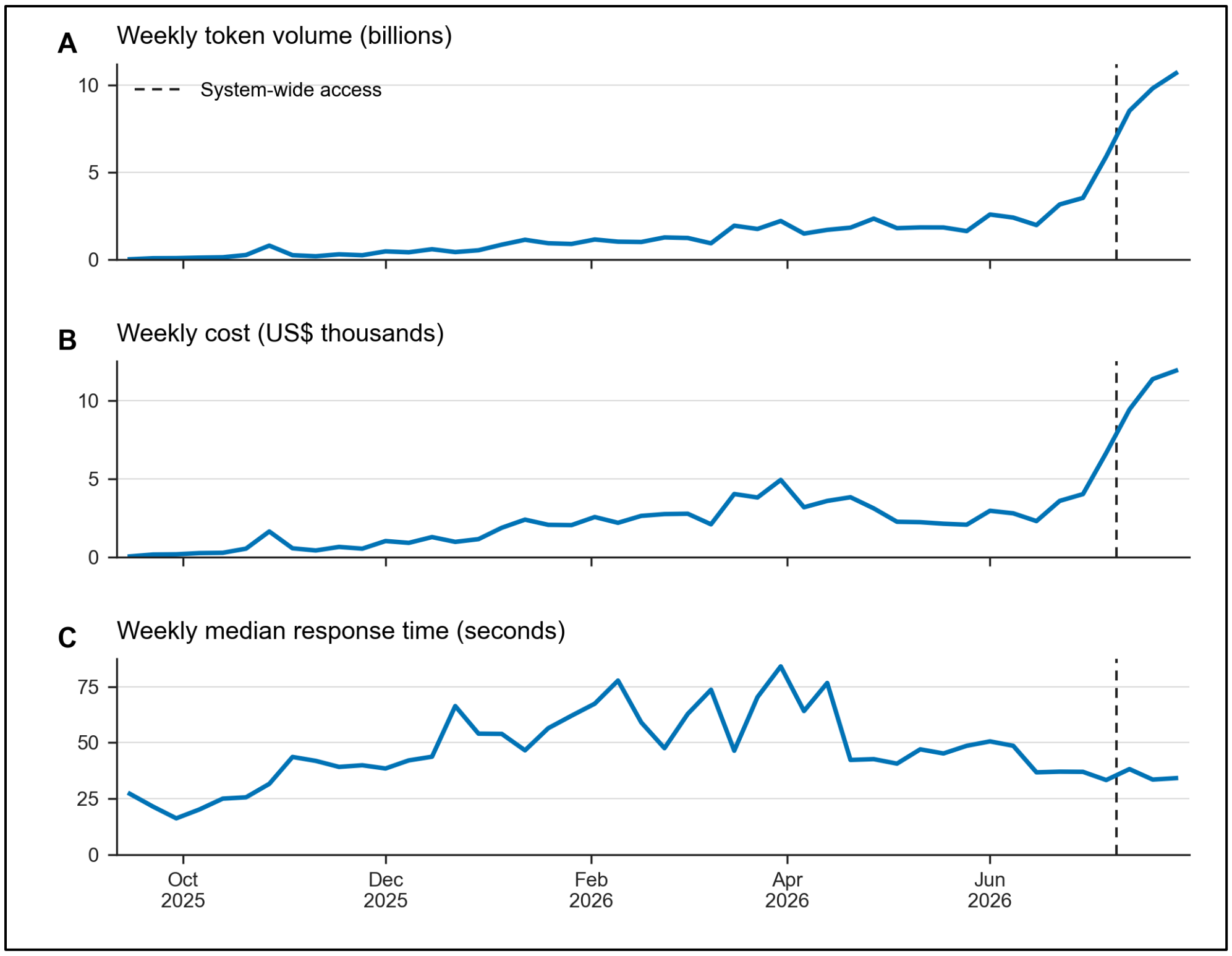

## Use cases and prompt design

Direct clinical care was the primary intended use for 92.5% of tasks. Summarization or synthesis was the most frequent operation (40.4%); retrieval, drafting or transformation, evaluation or derivation, and structured extraction each accounted for 10.9%-15.9%. Requested outputs were most often narrative summaries or timelines (40.5%), direct answers (21.8%), structured fields or tables (19.2%), and clinical documents or notes (14.5%; Figure 4). Table 5 reports event counts for every category. A distinct secondary use domain was explicit in 1,331 events (1.7%), most often education (953 events, or 1.2% of all events). Deidentified examples are provided in Tables S1 and S2 of the Supplementary Appendix.

Other prompts accounted for 54.4% of tasks, reused non-SmartTask prompts for 26.5%, and SmartTasks for 19.0%. When measured by users, reach was broader: 978 users (37.1%) used at least one SmartTask, 1,106 (42.0%) reused a non-SmartTask prompt across patients, and 2,457 (93.2%) used at least one other prompt; these groups overlapped. Every measured design feature was most prevalent in SmartTasks; multipart instructions occurred in 93.2% of SmartTasks compared with 65.8% of reused prompts and 31.7% of other prompts, and output specifications occurred in 94.2%, 71.4%, and 41.2%, respectively (Figure 4). Prompt sectioning occurred in 4.4%, 30.1%, and 35.3% of other, reused, and SmartTask events, and programmatic logic in 4.0%, 19.2%, and 25.3%. These differences describe prompt construction and do not establish prompt or response quality.

**Table 5. Prompt-defined use across 80,465 eligible task events.**

| Dimension | Category | Events, No. (%) |
|---|---|---|
| Primary use domain | Direct clinical care | 74,437 (92.5) |
| | Clinical research or registry | 1,714 (2.1) |

| Dimension | Category | Events, No. (%) |
|---|---|---|
| | Administrative or operational | 899 (1.1) |
| | Quality, safety, or compliance | 838 (1.0) |
| | Education | 117 (0.1) |
| | Not inferable | 2,460 (3.1) |
| Information operation | Summarize or synthesize | 32,474 (40.4) |
| | Retrieve | 12,775 (15.9) |
| | Draft or transform | 12,605 (15.7) |
| | Evaluate or derive | 11,794 (14.7) |
| | Extract or structure | 8,794 (10.9) |
| | Unclear | 2,023 (2.5) |
| Requested output form | Narrative summary or timeline | 32,619 (40.5) |
| | Direct answer | 17,552 (21.8) |
| | Structured fields, form, or table | 15,460 (19.2) |
| | Clinical document or note | 11,629 (14.5) |
| | Patient or external communication | 682 (0.8) |
| | Not specified or unclear | 2,523 (3.1) |
| Prompt-use mode | Other prompt | 43,797 (54.4) |
| | Reused non-SmartTask prompt | 21,346 (26.5) |
| | SmartTask | 15,322 (19.0) |

**Figure 4. Scout use cases and prompt-design features.** Panel A depicts the task-weighted distribution of intended-use domains, Panel B depicts the primary requested information operation, Panel C depicts the requested output format, and Panel D depicts prevalence of selected prompt-design features across prompt-use modes. Operations in Panel B are ordered by conceptual transformation of the record rather than prevalence. In Panel D, lines show the observed range across modes and are not uncertainty intervals. Prompt-design features were non-mutually exclusive; total may exceed 100%. SmartTask statuses were extracted from system metadata; reused non-SmartTask prompts were identical prompts applied across >= 2 distinct patient records. Semantic and structural labels were assigned to 46,485 unique prompts using GPT-5.6 Luna and mapped 80,465 eligible tasks. Prompt design features demote attributes rather than quality scores.

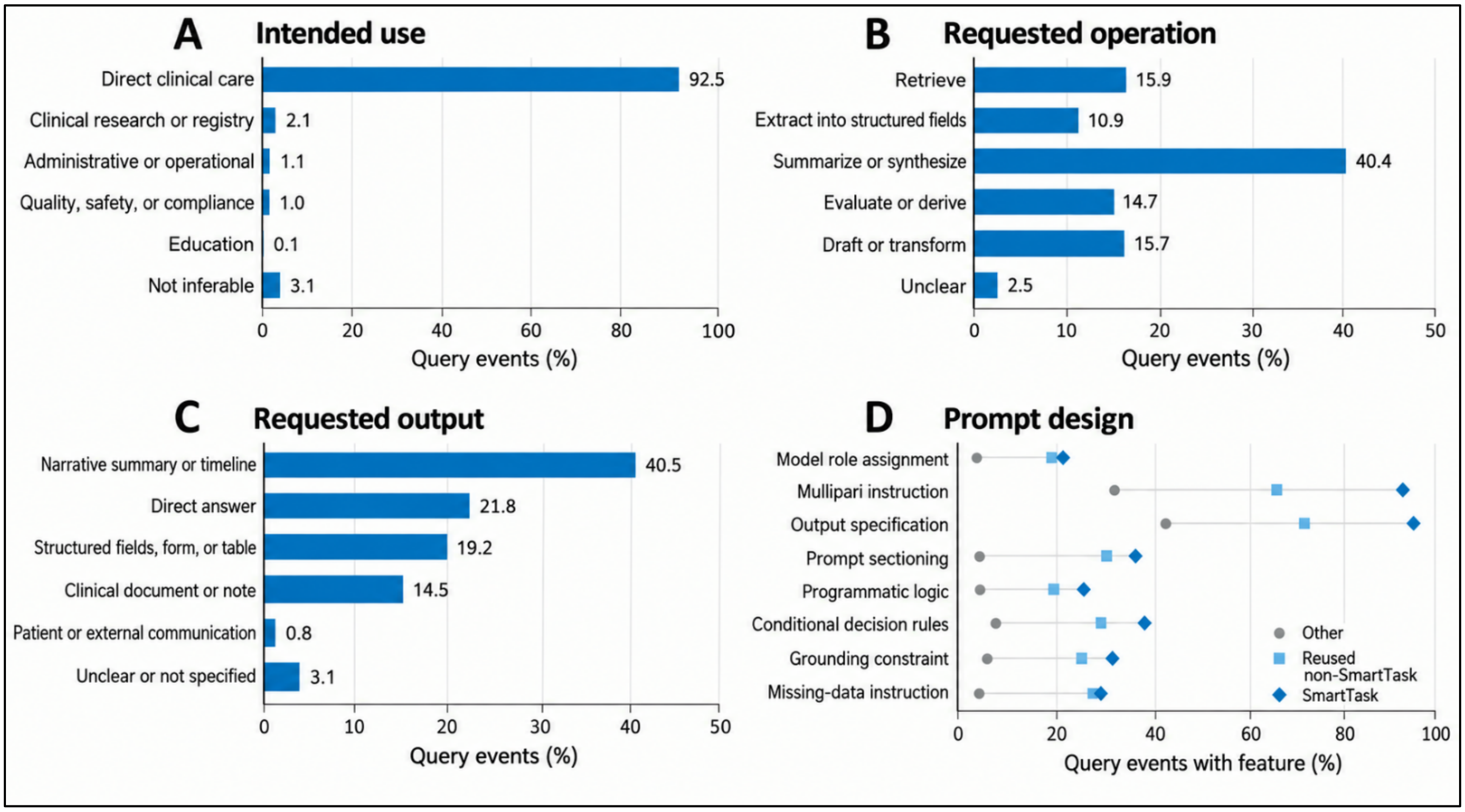

Requested tasks differed across intended-use domains (Figure 5). Within direct clinical care, summarization or synthesis was most frequent (43.3%). Research or registry tasks primarily requested evaluation or derivation (62.3%) or structured extraction (34.7%), whereas quality, safety, or compliance tasks primarily requested evaluation or derivation (83.8%). Administrative or operational tasks most often requested evaluation or derivation (38.7%) or drafting or transformation (26.3%). Structured output was requested for 92.7% of research or registry tasks and 77.8% of quality, safety, or compliance tasks, whereas narrative output was most frequent in direct clinical care (43.4%). SmartTasks accounted for 37.0% of quality, safety, or compliance tasks but 0.6% of research or registry tasks; reused non-SmartTask prompts accounted for 43.9% of research or registry tasks. These percentages describe the distribution of task events within each domain; education had only 117 events.

**Figure 5. Requested tasks and prompt deployment within intended-use domains.** Panel A presents the task-weighted distribution of requested information operations within each primary intended-use domain. Panel B displays requested output formats, and Panel C shows prompt-use modes. Values and heatmap shading represent the domain-specific task proportions; rows sum to 100% within each panel, subject to rounding. The not-inferable domain (2,460 events) was excluded. Unclear and unspecified output formats were collapsed. This analysis was derived from Scout task logs.

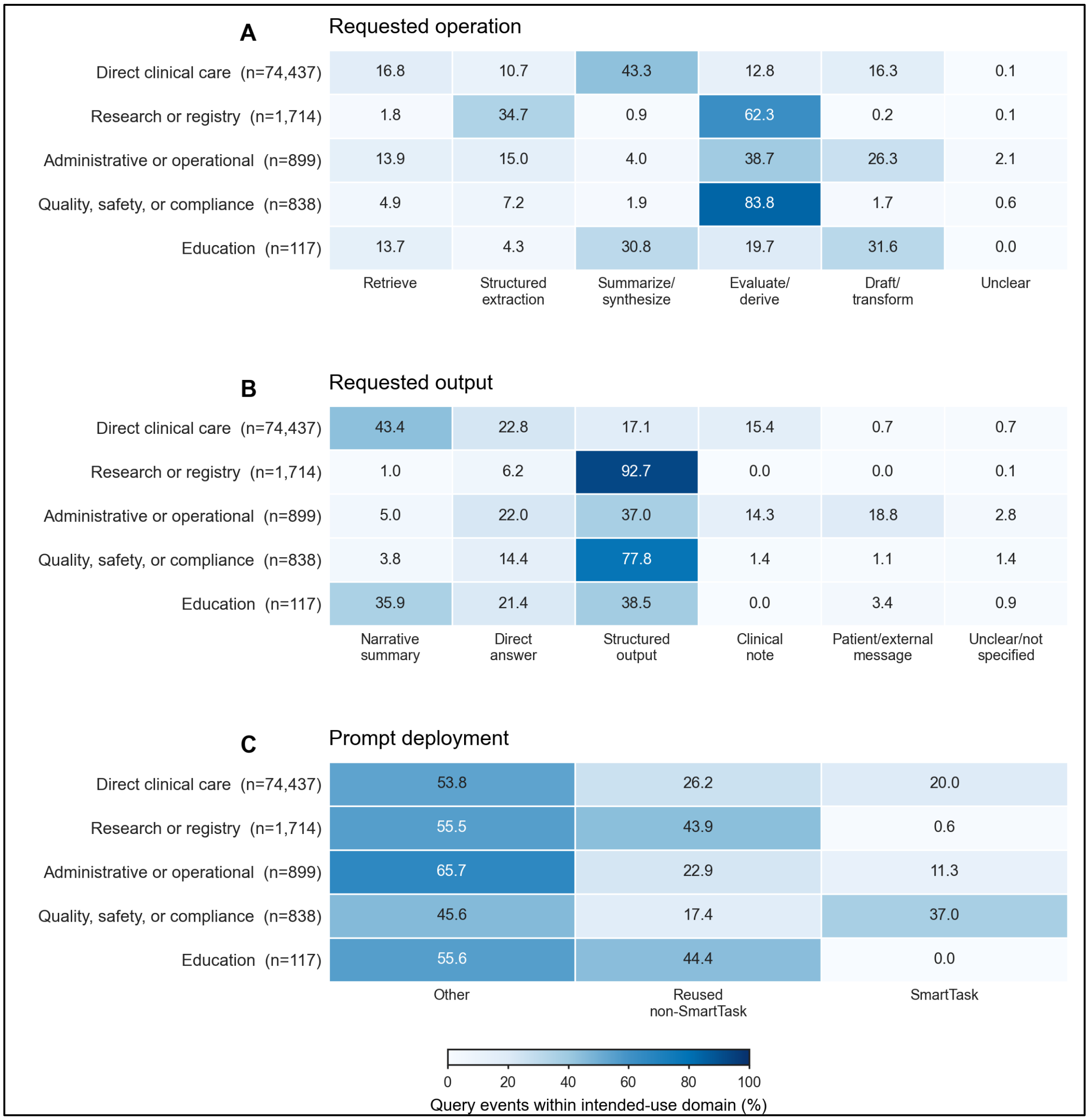

## Use by care practitioner role

Use varied across roles (Figure 6): attending physicians emphasized synthesis; residents and fellows had the greatest extraction and research shares; APPs had the greatest drafting and SmartTask shares; and registered nursing and other clinical staff more often requested evaluation. Associations were modest (Cramer's V, 0.15-0.18). The cohort comprised 30,657 attending events from 1,084 users, 22,691 resident or fellow events from 638 users, 14,564 APP events from 597 users, 5,349 registered-nursing events from 148 users, and 7,204 events from 168 other clinical staff; only 3 users had more than one recorded role, and each was assigned the modal role. Role-specific distributions are given in Table 6.

Direct clinical care accounted for 95.3% of attending, 89.9% of resident or fellow, 95.5% of APP, 83.2% of registered-nursing, and 89.8% of other clinical staff events. Residents and fellows had the largest research or registry share (6.2%); registered nursing had the largest quality or safety share (8.5%) and the largest administrative or operational share (5.1%). Attending physicians most often requested summarization or synthesis (50.6%). Residents and fellows had the largest structured-extraction share (20.2%), APPs the largest drafting share (21.5%), and registered nursing and other clinical staff the largest evaluation or derivation shares (43.4% and 22.0%). APPs had the greatest SmartTask share (32.7%), whereas other clinical staff had the greatest reused non-SmartTask share (38.8%).

Cost per task also varied by professional role. During the 29 days after system-wide access, median cost per task was $0.23 for registered nursing, $0.26 for physicians, and $0.35 for advanced practice providers; corresponding mean costs were $0.77, $1.10, and $1.43. Across the full observation period, registered nursing had the lowest mean cost per task among the three groups ($1.57), compared with physicians ($1.67) and advanced practice providers ($1.71). Registered nursing also had the highest observed task volume per user (36.1), compared with attending physicians (28.3) and advanced practice providers (24.4). Because registered nurses were not included in the July 2026 system-wide eligibility expansion and accessed Scout through trial and limited-launch pathways, this self-selected group may not be representative of the broader nursing workforce.

**Table 6. Use by care-practitioner role.**

| Role | Users | Events | Direct clinical care | Distinguishing pattern |
|---|---|---|---|---|
| Attending physicians | 1,084 | 30,657 | 95.3% | Summarization or synthesis was the most frequent operation within the role (50.6%) |
| Residents and fellows | 638 | 22,691 | 89.9% | Largest research or registry share (6.2%) and largest structured-extraction share (20.2%) |
| Advanced practice providers | 597 | 14,564 | 95.5% | Largest drafting or transformation share (21.5%); greatest SmartTask share (32.7%) |
| Registered nursing | 148 | 5,349 | 83.2% | Largest evaluation or derivation share (43.4%), quality or safety share (8.5%), and administrative or operational share (5.1%) |
| Other clinical staff | 168 | 7,204 | 89.8% | Evaluation or derivation 22.0%; greatest reused non-SmartTask share (38.8%) |

**Figure 6. Scout utilization patterns stratified by professional role.** Cell text indicates the percentage of role specific tasks across primary use domain (Panel A), information operation (Panel B), requested output formats (Panel C), and prompt-use modes (Panel D). Cell heatmap shading reflects the percentage-point deviation from the overall cohort share; gray is lower, white is comparable, and blue is higher. The scale is capped at +/-30 percentage-points. User were assigned their modal recorded role across deployment periods. APP denotes advanced practice provider. Analyses were exploratory and unadjusted for specialty, calendar period, or access duration.

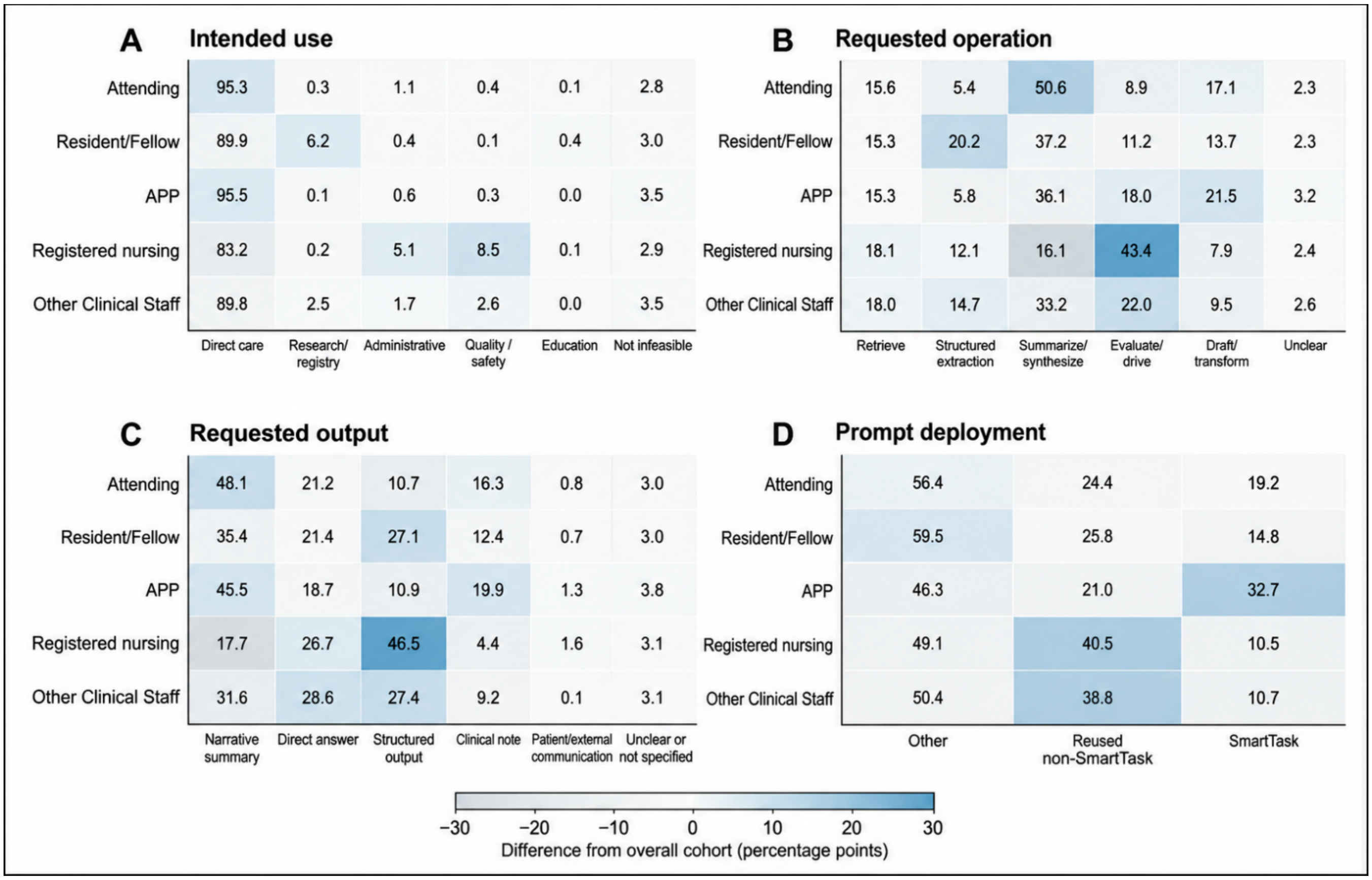

## User-reported experience

A follow-up subjective-experience survey completed by 100 users in May 2026 measured perceived workload, value, and willingness to recommend. On NASA Task Load Index scales scored out of 100, mean perceived performance was 72, mental demand 20, frustration 25, and effort 36. Mean ratings for value and for quality were each 4 of 5. Sixty-five respondents were promoters, 21 were passive, and 10 were detractors, giving a net promoter score of 57 among the 96 who answered that item. Sixty-six percent agreed that Scout would reduce their burnout, 22% neither agreed nor disagreed, 8% disagreed, and 4% did not respond. Among 84 substantive free-text comments, 93% described time savings, 82% quality improvement, 64% cost saving, and 37% revenue or capacity gains. Respondents described saving 5 to 15 minutes per patient on chart review and reported finding information they had missed on their own manual review.

*Source: May 2026 follow-up user survey. The survey was administered during the limited launch and describes that earlier-access population only. Responses were self-reported and were not linked to task logs.*

## Training completion and activation

Required training became available on May 1, 2026, and 3,258 care practitioners completed the course through September 22. By July 8, the day before system-wide access, 835 had completed it, equivalent to 10.1% of the 8,306 launch-eligible accounts; a further 1,836 completed it during the 29 days after system-wide access, bringing the total to 2,671 by August 6, or 32.2% of eligible accounts. Among those 2,671, 2,080 (77.9%) had used Scout by August 6 and 2,263 (84.7%) had used it by the September session extract. Median time from training completion to first use was 1.7 days (IQR, 1.3-5.4), and 22.5% of activating users began the same day they finished the course. In the first month of system-wide physician and APP access, uptake tracked training completion closely.

*Source: Scout learning-management-system completion records joined to the user-level session export on institutional user identifier. First use is drawn from that export, not the task-log cohort that serves as this manuscript's reference standard; the two differ by 101 users overall. Both extracts extend past August 6, 2026.*

## Provisioning, activation, and specialty distribution

To corroborate the task log-based adoption findings with independently sourced data, we obtained two user-level extracts. The first was a provisioning record for 833 users provisioned through trial and limited-launch pathways before system-wide Epic access. Of these, 815 carried a recorded provisioning date, spanning July 21, 2025, through July 1, 2026; for the remaining 18, a provisioning date was not recorded at the time of provisioning. The second was a session-level export of first Scout use with recorded specialty and care-practitioner title, containing 3,066 users. Of these, 2,534 first used Scout within the study window; 528 first used Scout after August 6, 2026, and 4 before September 10, 2025, and were therefore outside the analysis period. That 528 users initiated use in the month after the observation window closed indicates that uptake continued beyond the period analyzed here.

Of the 833 provisioned users, 707 (84.9%) had a recorded Scout session and 126 (15.1%) did not (Figure 7). Users were matched on exact user identifier; normalizing case or whitespace recovered no additional matches. Among the 707 activated users, provisioning-to-first-use lag could be calculated for 700, after excluding 6 users with no recorded provisioning date and 1 whose first session preceded the recorded provisioning date. Median lag was 6.6 days (IQR, 1.4-28.4 days). These estimates describe the earlier-access cohort specifically and were not extrapolated to the 8,306-account system-wide launch population, for which comparable individual-level provisioning dates were not available (see Discussion).

The session-level export's cumulative first-use curve (Figure 7 and, reproduced, Figure 1A) totaled 2,534 distinct users by August 6, 2026, compared with the 2,635-user task-eligible cohort analyzed throughout this manuscript. The difference is not evenly distributed across the study period. Truncation does not explain this gap because the export's earliest first-session timestamps precede September 10, 2025. Two extracts were generated at different times for different operational purposes, and the earliest access pathways were the least systematically recorded. The task-log cohort of 2,635 is the reference standard; the session export is reported because it independently reproduces the timing and shape of adoption, including the launch inflection.

By recorded primary specialty category, the largest represented groups among the 2,534 users who first used Scout within the study window were general internal medicine and hospital medicine (371 users, 14.6%), general outpatient or pediatric practice (266, 10.5%), anesthesiology and critical care (219, 8.6%), cardiology and vascular medicine (193, 7.6%), and psychiatry and behavioral health (141, 5.6%); the remaining users were distributed across 33 further specialty categories, 38 in total, none individually exceeding 5% of the cohort. A specialty category was recorded for all 2,534 users and a care-practitioner title for 2,489 (98.2%). This distribution is broadly consistent with the eligible-provider composition described in the Methods and does not suggest that adoption was concentrated in a small number of specialties.

**Figure 7. Provisioning-to-first-use activation lag among an earlier-access cohort.** Panel A shows the distribution of days from provisioning to first recorded Scout session among 700 users, drawn from the 833 users provisioned through trial and limited-launch pathways before system-wide Epic access. This excludes 126 users with no recorded session, 6 activated users with no recorded provisioning date, and 1 activated user whose first session preceded the recorded provisioning date. The distribution is capped at 120 days for display. Panel B shows activation status among all 833 provisioned users. These estimates are specific to this earlier-access cohort and are not a system-wide launch-cohort adoption rate. *Source: Authors' analysis of user-level provisioning and session records (provision_date.xlsx and users_spec_dept_firsts.csv).*

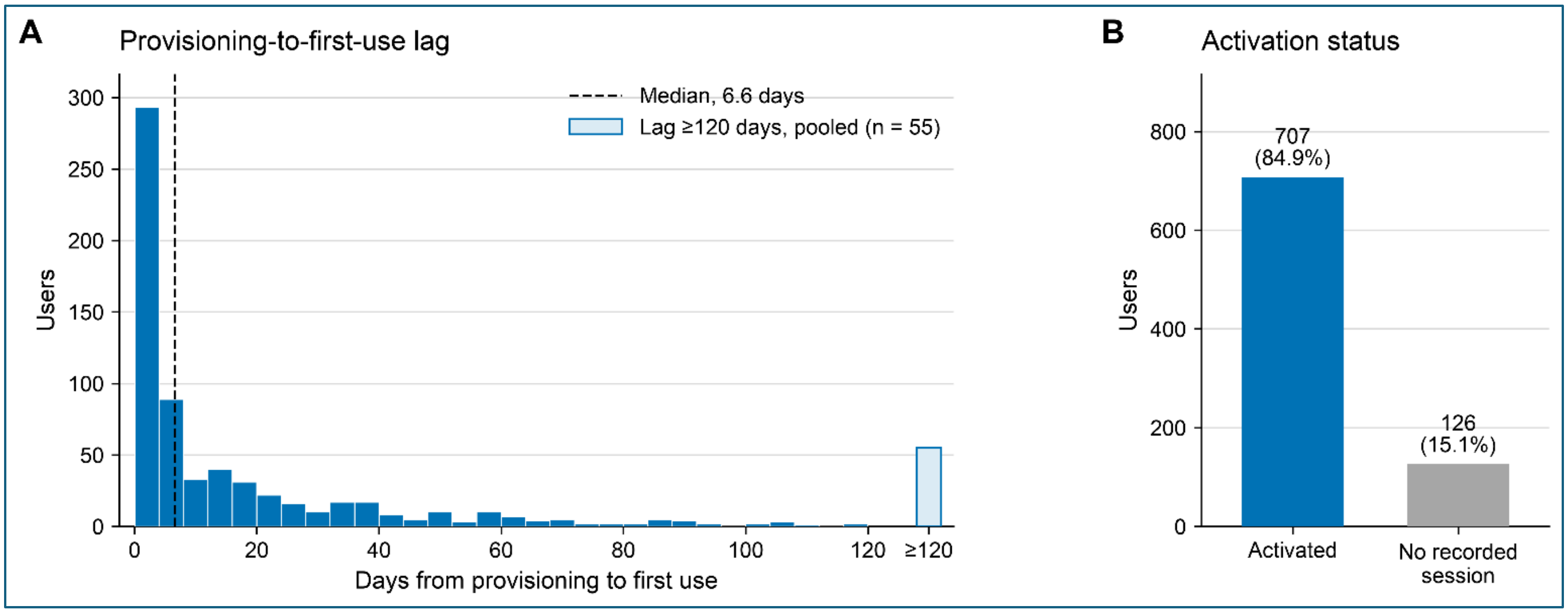

## Defects identified during staged expansion

During testing in May 2026, DIHI identified and corrected a rare failure mode in which a retrieval subagent could fail mid-task after beginning to return evidence, leaving Scout without supporting evidence at the moment it generated a final answer; because the evidence check ran only once, after the subagent's turn completed, this narrow timing gap could, in principle, allow an unsupported response, the failure Scout's citation-first design is meant to prevent. The evidence check was moved to run immediately before final-answer generation regardless of upstream failures, and two regression tests were added to confirm before every deployment that Scout cannot generate an answer with no supporting evidence. This incident also prompted development of the structured in-application feedback and error-reporting flow described below, replacing informal reporting channels that could not triage issues quickly enough as use grew.

Clinical task evaluation identified a separate medication-interpretation defect. In April 2026, clinicians evaluating a hospital medicine discharge summary task found that the model occasionally misinterpreted narrative dose descriptions and summed them incorrectly. Because the output looked plausible, only reviewers familiar with clinical documentation conventions could spot the error. We revised the task to reproduce medication information verbatim and to prohibit dose calculations. The defect was identified during task evaluation before clinical deployment; no affected output entered a patient’s chart. The correction was developed through May and implemented by June. It was incorporated first into the task under evaluation, then into the stored SmartTask, and finally into Scout’s system prompt, extending the safeguard to all users

### Governance during scale-up

SmartTask governance scaled from individual use to small-group evaluation and health-system approval, with increasing requirements for evaluation, education, and monitoring (Table 7). Personal SmartTasks required clinician review at each use; broader deployment required additional evaluation and oversight.

Each SmartTask had an identifier, and every invocation was logged with its response, citations, estimated cost, response time, and submitted feedback. These records supported evaluation across patients, reassessment after model changes, and identification of potentially affected users when defects emerged. Before system-wide access, required training and structured in-application error reporting were implemented.

By the end of the observation period, 752 SmartTasks had been invoked 15,322 times. Two had completed health-system approval; the remainder stayed at individual or small-group scale. The framework began operating during the study, and its pragmatic thresholds had not been independently validated.

**Table 7. Staged governance framework for reusable Scout SmartTasks.**

| Stage and scale | Evidence required to proceed | Oversight and status |
|---|---|---|
| Individual (1 user) | No additional evidence beyond the approvals authorizing Scout; the user validates the output for each use. | Self-governed |
| Trial (approximately 1-5 users) | Rubric-based assessment of accuracy, completeness, and relevance, with greater than 85% success in all three domains across at least 5 patient cases. | Local team |
| Limited launch (approximately 6-20 users) | At least 30 task runs evaluated for accuracy, completeness, relevance, and safety; successful evaluation verified; educational material prepared; results submitted to the Duke Institute for Health Innovation. | Local team with central review |
| Duke Approved (all eligible use-case users) | Ongoing automated fact-checking, cross-case consistency checks, and LLM-as-judge auditing, with major findings reported to governance. | Governance approved and centrally discoverable in the SmartTask library |

## Discussion

Scout was introduced to physicians and advanced practice providers through a randomized clinical trial, a limited launch, and subsequent Epic-integrated availability. Frontline clinicians specified task requirements and evaluated outputs, while senior leaders supported security review, established training requirements, and enabled access through Epic. Four implementation lessons emerged: clinical ownership of task design, traceability-based governance, centralized stewardship of shared infrastructure, and integration of training with product deployment.

**Clinical ownership.** Frontline clinicians are best positioned to author SmartTasks and assess their value and risk. Evaluation of a hospital-medicine discharge-summary task showed that medication assessments should preserve the practitioner’s wording and prohibit dose calculations. Incorporating this correction into Scout’s system design extended the benefit beyond the originating unit. Scout logs supported clinician review, measurement, and reporting to executive leaders. Patterns across 46,485 distinct prompts and 80,465 task invocations further showed that use varied more by professional role (Figure 6) than by care phase or data type. Task content primarily reflected specialty, requested operation (summarize, find, or compare), and chart time window. Because reviewers could not reliably infer care phase from task text, governance organized around care phases would have been impractical. Specialty-level expertise is therefore essential to task design and evaluation.

**Traceability-based governance.** Traceability may matter more than restricting prompt authorship. Open task authorship enabled unanticipated, specialty-specific questions, while restrictions shifted task development into personal documents outside governance. In the randomized trial, tasks developed without intensive optimization met noninferiority criteria for accuracy, completeness, and relevance and were superior for completeness.[7] Assigning identifiers to reusable SmartTasks and logging each invocation, including its response, citations, cost, and feedback—allowed sampling, re-evaluation after model changes, and identification of affected users when defects emerged. This created an auditable pathway from individual use to approval for broader use across clinicians. Of 752 SmartTasks invoked 15,322 times (Table 7), two completed centralized review for enterprise-wide use. We recommend open authorship with comprehensive logging, evidence-based expansion, review proportional to reach, and point-of-use error reporting.

**Central stewardship.** Centralized governance should secure, scale, and support the platform rather than prescribe local use. Senior leaders encouraged reducing operating costs. The implementation team removed the 'all history' and 'all encounters' shortcuts to encourage deliberate selection of query scope, while retaining whole-history searches through manually specified dates. A full history chart summary had a median cost of \$1.51 versus \$0.13 for a 30-day window, so an unnecessarily large time window resulted in excess costs. Central teams should own security, infrastructure, cost controls, training, support, and analytics for local evaluation and reporting.

**Training as part of the product.** Training functioned as an integral component of Scout's implementation rather than as an external prerequisite. Access was withheld until course completion. Among the 2,671 care practitioners who had completed training by August 6, 77.9% subsequently used Scout; the median interval from completion to first use was 1.7 days. This close temporal association suggests that readiness for enterprise deployment should be assessed across the integrated product, training, support, and access-provisioning pathway rather than by technical availability alone.

**Limitations.** This single-system study included only 29 days of follow-up after system-wide access. Implementation priorities precluded a consistent account-level adoption denominator and a complete incident series. Role and task analyses were exploratory, and cost estimates included only large language model API charges during a period that encompassed preoptimization development. Effects on safety, revenue, access to care, system-level cost, and quality were not established. Concurrent AI and quality-improvement initiatives will complicate attribution; future evaluations should prespecify outcomes and a consensus-based attribution strategy. These recommendations should be interpreted within the limits of a single-system experience.

## Declarations

**Funding:** This work was conducted and supported internally by the Duke Institute for Health Innovation, Duke Health. The study received no industry, federal, or other external funding, and the institutional sponsor had no role in study design, data analysis, interpretation, or the decision to submit for publication.

**Use of AI-assisted technologies in manuscript preparation:** OpenAI Codex, a GPT-based coding and writing assistant, was used to help organize source material, edit text, check internal consistency, and prepare Word review files. The authors reviewed and revised all AI-assisted material and take responsibility for the accuracy, integrity, originality, citations, analyses, and conclusions. AI-assisted tools were not listed as authors and were not cited as primary sources. GPT-5.6 Luna was used as a study instrument for exploratory prompt classification, as described in the Methods.

## Glossary of Terms

**Table G1. Glossary of terms used in this manuscript.**

| Term | What it means |
|---|---|
| Scout | An LLM-based search and synthesis tool, built and operated by the Duke Institute for Health Innovation (DIHI), that answers a care practitioner's natural-language question about one patient's record by drawing from notes, laboratory results, imaging, and other chart data and returning an answer in which every claim links back to its source. It does not diagnose or recommend treatment, and as deployed during the study period it did not retain memory between tasks. |
| ChatEHR | A comparable EHR-grounded large language model assistant developed at Stanford Medicine, reported in Shah et al. (Nature Medicine, 2026) and a companion usage-pattern study.[(6)] ChatEHR is the closest published analogue to Scout and is used in this manuscript as an external point of comparison for adoption speed and use patterns. |
| Maestro Care | Duke Health's name for its Epic electronic health record environment. |
| Limited launch | Scout's entire non-trial access period before official launch, September 22, 2025 through July 8, 2026. Access was granted under the research protocol: first to experienced DIHI collaborators for interface and user-experience feedback, then to specialty groups for feedback, and from there by word of mouth. Distinct from the Limited Launch stage of the SmartTask governance ladder (Table 7), which describes how a single reusable prompt scales. |
| Scout task | A natural-language instruction a user writes describing what they want from one patient's record. Often a directive rather than a question, which is why DIHI uses task in preference to query or question. The text of the instruction is its prompt; a task saved for reuse is a SmartTask. |
| SmartTask | A prompt that a Scout user has saved to the Scout library so it can be recalled later by typing a period followed by the task's shortcut name, instead of retyping or pasting it from elsewhere. Some SmartTasks are authored and used by individual |

| Term | What it means |
| --- | --- |
| | care practitioners; others progress through DIHI's governance ladder (Table 7) to become centrally vetted, health-system-approved tasks. |
| DIHI | The Duke Institute for Health Innovation, the team that built and operates Scout and that manages its evaluation, governance, training, and user support. |
| Epic security and access provisioning | The Epic infrastructure, enterprise IT security review, and system-wide access-provisioning functions through which eligible users obtain Scout access after completing required training. |
| ABCDS | Duke Health's Algorithm-Based Clinical Decision Support Oversight Committee, which provides clinical governance review for algorithmic tools used in care. |
| Evidence card | The panel Scout displays alongside a citation, showing the original note or data point a claim was drawn from, so a user can verify it without leaving the tool. |
| LLM-as-judge | An automated evaluation method in which a separate large language model reviews another model's output for factual support against source material. Scout's evaluation program uses this as a first-pass screen, always followed by human adjudication before a finding is reported as fact. |

*Source: Authors' summary, condensed from internal program documentation.*

# Supplementary Appendix: Real-World Adoption, Use, and Governance of an LLM-Based Electronic Health Record Search and Synthesis Platform

## Supplementary Methods

### Data preprocessing and quality controls

The source export was a UTF-16 tab-delimited file with 214 columns. The preprocessing pipeline created a UTF-8 comma-delimited analytic copy while retaining the source fields and appending derived variables and quality flags. The canonical event time was taken from session_start_at, with task Datetime (str) and Date (p) used sequentially as fallbacks. The analytic user identifier used tableau_matching_netid when present and NetID otherwise. A pseudonymous user key was generated for downstream analyses.

A repeated event record was defined by the combination of analytic user identifier, session identifier, event identifier, canonical timestamp, and prompt text. A record was flagged for suspected schema misalignment when at least two of the following signals were present: invalid demo-mode value, invalid feedback value, invalid manuscript-period value, invalid numeric citation count, invalid numeric session duration, or disagreement between specialty sources.

The cleaned dataset retained direct identifiers and was therefore treated as a protected analytic source. Prompt text was retrieved from the protected source only for classification. The classifier received the prompt text but not the patient chart, model response, raw medical record number, or raw user identifier. The event-level analytic classification file excluded prompt text, raw patient and user identifiers, SmartTask identifiers, responses, and free-text classifier rationales.

### Derived quantitative measures

#### Retention

Monthly cohort retention was period-specific rather than continuous. For a cohort beginning in calendar month c, the month-m numerator was the number of cohort members with one or more eligible events in calendar month c+m, and the denominator was the number of users in the cohort. A user did not need to have activity in all intervening months. Month 0 was 100% by definition. The August 2026 first-use cohort and all August follow-up cells were excluded because the observation period ended on August 6. Blank cells represented unavailable follow-up and were not imputed.

#### Power-user classification

Tasks per active day were calculated as total eligible tasks divided by the number of distinct dates with an eligible task. Exactly the highest-ranked quintile was assigned to the power-user group. Ties at the quintile boundary were resolved deterministically using total tasks and then active days as secondary ordering variables. Tokens per observed task used only tasks with recorded token values. User-level token-per-task distributions were compared using a two-sided Mann-Whitney U test.

#### Weekly resource measures

Calendar weeks began on Monday. A week was complete only if all seven days fell within the observed event period. Estimated inference cost was calculated using Azure OpenAI API pricing for the GPT-family models powering Scout and excluded other infrastructure and program costs. Token and cost totals were not adjusted for the number of users or events. Response time in seconds was the cleaned session duration multiplied by 60 among events flagged as having a response. The weekly median was the primary response-time measure; the 90th percentile was retained as a descriptive tail measure.

## Prompt-use mode

Prompt-use mode was assigned hierarchically at the event level. A SmartTask is a reusable prompt that a Scout user has saved to the Scout library and recalls by typing a period followed by the task's shortcut name. Analogous to EHR dot phrases, SmartTasks could be invoked across patient records but instructed Scout to search and synthesize record data. SmartTasks vary in provenance. Many are written and saved by an individual care practitioner for their own use and have not undergone formal development or validation; these remain at individual scale, because unit-wide and enterprise-wide use requires progression through the staged governance ladder in Table 7, which is where co-development and validation with Duke Institute for Health Innovation technical and clinical teams occurs. Because SmartTask status was determined from system metadata, the analyses reported here do not distinguish individually authored from co-developed SmartTasks. SmartTask status required a nonempty SmartTask identifier; blank strings, missing values, and empty lists did not qualify. Among non-SmartTask events, exact observed reuse required identical prompt text in at least two distinct patient records. All remaining events were classified as other, including structured prompts observed in only one patient record and exact repetitions confined to a single patient record. Because the rule was event-level, the same prompt observed in both SmartTask and non-SmartTask contexts could contribute SmartTask events and, when the non-SmartTask observations spanned at least two patient records, reused non-SmartTask events.

## Semantic classification codebook

The semantic framework separated intended use from the task requested of Scout. Use domain captured why Scout was being used; information operation captured what Scout was asked to do; and output form captured the desired product. These axes described distinct dimensions rather than competing categories. For example, a prompt with a research intent could request an eligibility evaluation and ask for the result as either a direct answer or a structured table. The semantic classifier characterized the requested terminal deliverable rather than every intermediate instruction. Each axis was coded independently and was mutually exclusive within that axis: each prompt received one primary information operation, one requested output form, and one primary use domain. A secondary use domain was assigned only when the prompt explicitly requested a second, substantively different purpose.

### Primary information operation

- **Retrieve:** one or a few directly documented facts, dates, results, notes, or events.
- **Extract or structure:** a predefined multi-field abstraction, form, checklist, or table without a substantive judgment.
- **Summarize or synthesize:** an integrated narrative, history, explanation, or timeline.
- **Evaluate or derive:** a calculation, stage, score, eligibility or adherence determination, risk assessment, gap analysis, or recommendation.
- **Draft or transform:** a clinical note, history of present illness, discharge section, letter, message, or rewritten communication.
- **Unclear:** no usable terminal operation could be determined.

When multiple operations were present, the terminal product took precedence: document or communication drafting, then derived judgment, predefined field extraction, narrative synthesis, and direct fact retrieval.

### Requested output form

Output form was classified as a direct answer; structured fields, form, or table; narrative summary or timeline; clinical document or note; patient or external communication; not specified; or unclear. "Not specified" indicated that the task was interpretable but did not name an answer form, whereas "unclear" indicated that the requested product could not be determined.

### Primary and secondary use domain

The primary domain represented the terminal purpose and was classified as direct clinical care; clinical research or registry; administrative or operational work; quality, safety, or compliance; education; or not inferable. A secondary domain was assigned only when a second, different, substantive deliverable was explicit. Use of a patient record alone did not create a secondary direct-clinical-care label. Named programs or networks alone did not establish research use. Patient-specific retrieval, synthesis, assessment, care planning, treatment interpretation, and clinical documentation defaulted to direct clinical care unless the prompt explicitly established another terminal purpose.

## Prompt-design classification codebook

Prompt design was classified independently of semantic use and observed reuse. Features were coded only when explicitly present. Prompt length, apparent clinical complexity, and likely user intent were not used as proxies. Features could coexist and were not combined into a sophistication or quality score. The design features were independently coded and non-mutually exclusive: one prompt could contain any number of positive features. Consequently, feature percentages were calculated separately and were not expected to sum to 100%.

### Model role assignment

Model role assignment was classified as none, generic role, or domain-expert role. A positive label required a direct instruction assigning the model a role, such as "you are" or "act as." A user's self-description, intended audience, or requested document type did not constitute model role assignment.

### Directive form

Directive form was classified as a keyword fragment, a single instruction or question, or a multipart instruction. Multipart instructions required at least two separately actionable requirements, deliverables, constraints, or conditional behaviors. Length alone did not make a prompt multipart.

### Input specification

Input specification used the precedence order input-field schema, named inputs, and none. Named inputs required an explicit source or supplied object, such as nursing notes, medication administration records, laboratory results, imaging reports, or pasted text. An input-field schema required labeled parameters, structured variables, or an input object. Implicit access to the patient chart was not an input specification. Information requested in the answer was treated as an output requirement rather than an input.

### Output specification

Output specification used the precedence order machine-readable schema, literal output template, field specification, named output form, and none. A named output form identified an artifact such as a summary, table, note, letter, message, or timeline. A field specification enumerated required fields or sections. A literal template supplied an actual response skeleton or placeholders. A machine-readable schema required JSON, YAML, XML, or another structured object. The operation "summarize," without a specified artifact or organization, did not itself constitute an output specification.

### Prompt sectioning and programmatic logic

Prompt sectioning required visible organization of the instruction using headings, labeled blocks, or multiple separated requirement sections. A request for a sectioned answer did not make the prompt itself sectioned. Programmatic logic required reusable, configuration-like control architecture containing at least two of the following: named variables or fields, explicit source-to-field mappings, controlled output codes, repeated or nested branches, or template-rendering rules. A single clinical threshold or isolated if/then statement could be coded as a conditional decision rule without being programmatic. Sectioning, programmatic logic, and conditional decision rules were coded independently and could coexist.

### Exemplars

Exemplars were classified as none, a completed format example, or a worked input-output example. Blank templates, placeholders, and fill-in skeletons were literal output templates rather than exemplars; a populated demonstration was required.

### Process instructions

Three independent process features were coded. A rationale request required an explicit request for explanation, justification, supporting basis, or reasons. An ordered analytical procedure required the model to perform analytical or retrieval steps in a specified order; an ordered answer layout did not qualify. Conditional decision rules required explicit if/then rules, thresholds, decision trees, eligibility logic, stopping rules, or behavior contingent on a finding.

### Safeguards and scope

A grounding constraint explicitly restricted the model to chart or supplied evidence or prohibited unsupported inference, speculation, or fabrication. A missing-data instruction specified how absent, unknown, unavailable, or undocumented information should be represented. Prompt-embedded scope required an explicit time window, encounter, admission, source subset, or other record boundary in the prompt text. The system-selected encounter range was not used to infer prompt-embedded scope.

### Explicit constraints

Five independent constraint types were coded: length or quantity; ordering; presentation format; exclusion of specified content; and audience, tone, or reading level. A named output artifact was not counted again as a presentation constraint unless additional styling was explicit. Grounding language was not counted again as an exclusion constraint unless it separately excluded output content.

### Review flag

The operational review flag was reserved for empty, corrupted, malformed, non-task, test, or genuinely uninterpretable prompts. It was not used as a research outcome.

## LLM implementation and quality assurance

The semantic and design classifiers were run separately with GPT-5.6 Luna at medium reasoning effort. Each call returned a predefined JSON object with enumerated categorical values and JSON booleans. Responses were accepted only after schema validation. Successful results were cached by prompt, model, reasoning effort, codebook version, schema, and corpus manifest. Free-text rationales were retained only in protected caches and were excluded from analytic outputs. Full-corpus processing required complete one-to-one reconciliation of both classification passes with the distinct-prompt manifest and application of the resulting labels to all corresponding task events.

The exploratory classification framework and codebooks were developed and clinically validated by a physician investigator on the study team. Development used a simple random sample of 500 unique eligible prompts to refine the clinical definitions, category labels, examples, exclusions, and ambiguous boundaries. The development sample was not used as a prevalence estimate.

After the codebooks were frozen, the physician investigator independently reviewed a separate sample and subsequently examined disagreements and boundary cases with the analytic team. This physician-led process assessed the framework's clinical relevance, interpretability, and consistency of application and established face and content validity for an exploratory classification. It was not a formal criterion-standard accuracy study; sensitivity, specificity, and overall classifier accuracy were therefore not estimated. Four candidate semantic axes-requested content scope, explicit value target, context sufficiency, and explicit workflow stage-were not carried forward because physician review found their classification quality insufficient for reliable interpretation from prompt text. The complete frozen classifier instructions are reproduced below.

## Care-practitioner role analysis

Users were classified as attending physicians, residents or fellows, advanced practice providers (APPs; physician assistants, nurse practitioners, certified registered nurse anesthetists, and certified nurse midwives), registered nursing, or other clinical staff. Recorded care-practitioner title was used to resolve conflicts between legacy APP and nursing labels: physician assistants, nurse practitioners, certified registered nurse anesthetists, and certified nurse midwives were assigned to the APP group, whereas registered nurses, graduate registered nurses, and clinical nurse specialists were assigned to registered nursing. When a detailed title was unavailable, the existing role category was retained. When a user had more than one recorded role, the role associated with the largest number of that user's events was assigned; ties were resolved deterministically by role label. Role analyses pooled all eligible events across deployment periods. The analytic role groups described observed users rather than eligibility for system-wide access; the registered-nursing group therefore reflected recorded roles across the full observation period and all access pathways, including users provisioned through trial or limited-launch pathways; it was not a July eligibility category. For each outcome, an event-count contingency table crossed the five role groups with the prespecified categories. Cramer's V was computed from this table. For each of 10,000 permutations, the modal role labels were randomly reassigned across users and each user's entire vector of event counts was retained. The permutation P value was (b+1)/(10,000+1), where b was the number of permuted statistics greater than or equal to the observed statistic. A fixed seed was used for reproducibility.

Analyses were conducted in Python using pandas, NumPy, and Matplotlib.

## Exploratory cross-dimensional description

To distinguish why Scout was used from what it was asked to produce and how the request was deployed, we cross-tabulated primary intended-use domain with requested information operation, requested output form, and prompt-use mode. Percentages were task weighted and calculated within each domain. The not-inferable domain was retained in the complete marginal distribution but omitted from the visualization because a domain-specific task distribution was not interpretable. Unclear operations were retained; unclear and not-specified output forms were combined for display. This analysis was descriptive.

## Implementation chronology and governance abstraction

The implementation chronology was assembled from contemporaneous program records and confirmed by the implementation team. Trial enrollment began in September 2025 and data collection continued through the end of that year. From September 22 through December 31, care practitioners and health care administrators with experience on funded Duke Institute for Health Innovation projects were also invited to use Scout and provide interface feedback. This early period included 213 users: 82 faculty physicians, 48 residents and postgraduate trainees, 15 advanced practice providers, 23 nurses, and 45 other clinical staff, including pharmacists and physical therapists.[7] These users submitted 6,641 tasks, and use reached approximately 70 weekly active users.

Formal Duke Health Information Security Office review occurred from January through April 2026, with security approval on April 3. Confirmation from Duke Compliance and Algorithm-Based Clinical Decision Support governance leadership followed on April 27. During this interval, referral-review and research-screening extensions at the Duke Cancer Center added 22 users; extensions of trial enrollment added 40 psychiatry and behavioral health users, 16 transplant center users, 14 cardiology users, 6 nephrology users, and 12 medical genetics and genomics users. Growth from April through early July was driven principally by word of mouth while consent, training, and collection of user role and interest information continued. The first system-wide learning-management-system training version was circulated on June 11. Scout was enabled in Maestro Care on July 8 and made available to physicians and advanced practice providers on July 9. A contemporaneous July 7 go-live briefing documented a launch-eligible population of 8,306 provider accounts within the Duke Epic service area; Community Connect and other external service areas were excluded. Eligible provider types comprised physicians and residents, including anesthesiology and osteopathic medicine; advanced practice providers (APPs), including physician assistants, nurse practitioners, certified registered nurse anesthetists, and certified nurse midwives; and dental, podiatric, and chiropractic care practitioners. Registered nurses were not eligible for system-wide access. Users provisioned through earlier trial or limited-launch pathways retained access in the longitudinal cohort, including registered nurses and other staff outside the system-wide eligibility group. Eligibility did not confer access by default: Scout remained unavailable until the provider completed the required tracked learning-management-system training, after which access was provisioned through Epic. Weekly quantitative analyses used July 6 only as the Monday beginning the calendar week that contained system-wide access.

Daily use initiation was calculated separately before and after system-wide Epic access. First use was the date of a user's earliest eligible task, not the date on which access was provisioned. The pre-access mean was the number of users whose first eligible task occurred from September 10, 2025, through July 8, 2026, divided by the 302 calendar days in that inclusive interval. The early system-wide-access mean used July 9 through August 6, divided by 29 calendar days. The latter was also divided by the 8,306 provider accounts in the launch-eligible population. This percentage contextualized initiation against that eligible population; it was not an account-level adoption rate because the numerator was not restricted to the 8,306-account launch group and account-level provisioning dates were unavailable for that period. Training-completion records were available and are used separately for the activation analysis reported in the Results.

Governance stages were abstracted according to the number of intended users, the evidence required before expansion, and the responsible oversight level. The stages were Individual, Trial, Limited Launch, and Duke Approved. These SmartTask governance stages are named independently of Scout's own limited-launch deployment period described in the Methods. The framework applied to reusable SmartTasks rather than to every ad hoc task. The requirements described in Table 7 are implementation standards; this study did not independently test whether an individual requirement reduced errors or improved patient outcomes.

Incident reporting pathways were described before and after preparation for system-wide access. Trial period issue reports could enter through in-application feedback, direct email, or a Microsoft Teams support channel. Before broader launch, the implementation team created a structured in-application reporting flow designed to reduce the risk of delayed or misdirected triage. The interface enabled users to submit feedback or report errors at the point of use; each submission triggered notification to the technical team responsible for triage and resolution. Training requirements were similarly described as implementation controls. Completion was tracked for the system-wide learning-management-system video, and access through Maestro Care was withheld until completion. Earlier trial access also required consent, policy acknowledgment and attestation, and IRB approval. These process descriptions were not analyzed as comparative interventions.

## Illustrative Deidentified Prompt Examples

Tables S1 and S2 provide representative examples selected from actual Scout prompts after classification. No synthetic prompts were substituted. Direct and potentially identifying information was removed or replaced with bracketed descriptors. Long prompts were abridged and capitalization or punctuation was lightly standardized without changing the operative instruction. Examples were selected to clarify category boundaries rather than to represent the most frequent prompts or recommended prompting practice.

**Table S1. Deidentified examples from actual Scout prompts for the semantic categories displayed in Figure 4 of the main article.**

| Axis | Category, events, No. (%) | Actual prompt examples |
|---|---|---|
| Use domain | Direct clinical care, 74,437 (92.5) | 1. "summarize medical history as it relates to sleep medicine" 2. "Please summarize this patient's sleep history and reason for referral to come see us in the sleep clinic succinctly." |
| Use domain | Clinical research or registry, 1,714 (2.1) | 1. "Looking at the attached inclusion and exclusion criteria, does this patient meet the screening requirements for enrollment into this clinical trial?" 2. "Using the attached reference document, tell me if this patient is eligible for this clinical trial." |
| Use domain | Administrative or operational, 899 (1.1) | 1. "Can you provide a report detailing each PT encounter billing documentation flowsheet data. Please highlight any time that a charge is selected for less than 8 minutes?" 2. "Is there currently an open [CARE-MANAGEMENT PROGRAM] case for this patient? Confirm whether a case is already active before proceeding with outreach. If a case is open, provide the case type, assigned staff, and open date if available." |
| Use domain | Quality, safety, or compliance, 838 (1.0) | 1. "What is the CMS time zero for this patient meeting CMS SEP-1 criteria? Please provide the exact date and time." 2. "Find any notes that are missing any of the following: [REQUIRED DOCUMENTATION ELEMENTS]; provide the date of the note if a note is missing this information." |
| Use domain | Education, 117 (0.1) | 1. "Review the patient's case and identify three high-yield teaching points for residents or medical students, including evidence-based management pearls, relevant guidelines, and common pitfalls." 2. "summarize the delivery course and what i need to know as a NICU resident" |
| Use domain | Not inferable, 2,460 (3.1) | Examples not shown because these prompts were generally context dependent, opaque, or non-task-like. |
| Information operation | Retrieve, 12,775 (15.9) | 1. "When this patient had a nasogastric tube, did they have a clamp trial or trickle tube feeds prior to removal of their nasogastric tube?" 2. "Why was this patient referred to gastroenterology recently?" |
| Information operation | Extract or structure, 8,794 (10.9) | 1. "Extract: highest/lowest temperature, lowest SBP, vasopressor use, invasive mechanical ventilation, cardiac arrest, and worst mental status. Use the worst documented value/status, include evidence and |

| Axis | Category, events, No. (%) | Actual prompt examples |
| --- | --- | --- |
| | | date/time, return Unknown if unavailable." 2. "Please provide just the summary table for the following information: reason for admission/mechanism of injury; GCS on arrival-provide a number within [VALID RANGE]; was there a headstrike-yes or no; was there loss of consciousness-yes or no; results of head/brain CT or MRI; was there change in mental status-yes or no." |
| Information operation | Summarize or synthesize, 32,474 (40.4) | 1. "Please summarize this patient's sleep history and reason for referral to come see us in the sleep clinic succinctly." 2. "Provide a detailed summary of the patient's complete medical history including: past medical history, surgical history, family history, social history, allergies, current medications, and a comprehensive review of systems. Organize the information chronologically and highlight any significant patterns or relationships between conditions." |
| Information operation | Evaluate or derive, 11,794 (14.7) | 1. "Based on this patient's initial vital signs, triage note, history, and medications, can you determine the correct ESI level?" 2. "Review medication adherence patterns and identify potential compliance issues." |
| Information operation | Draft or transform, 12,605 (15.7) | 1. "I am the consulting cardiologist. Another cardiologist will take over for me tomorrow, and I am emailing him a signout report on each of my patients. Help me write a brief, few sentence signout report for him." 2. "Create a comprehensive discharge summary including: primary and secondary diagnoses with ICD codes, hospital course summary, procedures performed, significant findings, discharge medications with reconciliation notes, discharge instructions, follow-up appointments needed, pending results to follow up, and specific red flags that should prompt return to ED." |
| Information operation | Unclear, 2,023 (2.5) | Examples not shown because the terminal operation could not be determined. |
| Output form | Direct answer, 17,552 (21.8) | 1. "is patient on immunotherapy?" 2. "Why is this patient having surgery today?" |
| Output form | Structured fields, form, or table, 15,460 (19.2) | 1. "Please look up and summarize in table format most recent ECGs, echocardiograms, cardiac stress tests, heart catheterizations (left, right, and any interventions), PFTs, or chest xrays. Also include any clinically relevant imaging related to history of present illness." 2. "Summarize the assessment of the most recent care/evaluations from all specialists the patient sees. Include the specialty, date of most recent visit, provider name, summary of visit and recommendations, and return-to-clinic instructions. Use the following format: Specialty: Provider: Summary of visit: Plan:" |
| Output form | Narrative summary or timeline, 32,619 (40.5) | 1. "summarize the patients surgical history" 2. "Give a summary/timeline/overview of this patient's psychiatric history/journey." |
| Output form | Clinical document or note, 11,629 (14.5) | 1. "Concisely summarize this patient's past medical history and current encounter in the format of an HPI." 2. |

| Axis | Category, events, No. (%) | Actual prompt examples |
|---|---|---|
| | | "Please write an overnight intensive care unit assessment and plan utilizing critical care diagnoses and any relevant major comorbid conditions." |
| Output form | Patient or external communication, 682 (0.8) | 1. "Create a concise patient friendly review of lab results that focuses on the overall big picture from each result. For results with multiple components provide an overall summary of the test and only expand on abnormal elements of clinical significance. Look at results from the most recent encounter." 2. "Write a medical exception to school testing based on the patient's documented diagnoses, conditions, disabilities." |

*Source: Authors' analysis of deidentified Scout prompts.*

The not-specified and unclear output-form categories remained in the complete distribution (544 [0.7%] and 1,979 [2.5%] events, respectively) but were omitted from the example table because reproducing them added little interpretive value.

**Table S2. Deidentified examples from actual Scout prompts for the non-mutually-exclusive prompt-design features displayed in Figure 4 of the main article.**

| Feature, events, No. (%) | Actual prompt examples |
|---|---|
| Model role assignment, 8,142 (10.1) | 1. "You are a shrewd internal medicine physician preparing this patient for surgery. Summarize the patient's past medical history with attention to perioperative risks. Keep the summary concise." 2. "You are an experienced hospitalist creating an attending-to-attending handoff draft. Write a single paragraph and briefly summarize why the patient is hospitalized, their current clinical trajectory, major active problems, important diagnostic findings, key management decisions, anticipated next steps, discharge barriers, and any issues requiring close follow-up by the next attending." |
| Multipart instruction, 42,234 (52.5) | 1. "Generate an up-to-date problem list for this patient. Organize all active medical problems by severity/acuity (critical, high, moderate, low). For each problem, include date of onset, current status, relevant recent labs/imaging, current treatment plan, and any pending workup or consultations. Also note any resolved problems from this admission." 2. "I want you to write a pre-operative assessment note for the [PREOPERATIVE CLINIC] for her procedure [PROCEDURE]. Please start with HPI and make it clear and concise. Then PMH-all diseases, when diagnosed, status, treatment, relevant surgeries, testing, and procedures associated with the disease. Include the last hospitalization, surgical history, recent laboratory testing, cardiac testing, and anesthesia history. Do not include medications or allergies." |
| Output specification, 47,712 (59.3) | 1. "Summarize osteoporosis history including risk factors, fracture history, and treatment history. Output should be paragraph format for an e-communication note." 2. "Provide oncology history for a genetic counseling clinic note. Include dates of procedures and treatments. If there are multiple cancers, organize information by primary cancer type. List cancer history in reverse chronological order. Do NOT provide summary at the end. Do NOT repeat information." |
| Prompt sectioning, 13,763 (17.1) | 1. "#Role: You are a clinical provider responsible for the comprehensive care of a hospital patient. #objective: Sythesie information pertinent to this patient's current hospitalization and create a very brief overview that illustrates the admission reason and main active issues. #Structure: Make this brief; short-hand writing style |

| Feature, events, No. (%) | Actual prompt examples |
|---|---|
| | is acceptable and does not need to be in complete sentences." 2. "Briefly summarize the patient's current clinical context and illness trajectory as framed during the discussion. Note the patient's and/or family's stated level of understanding of the prognosis. STYLE RULES: Third person throughout; single flowing narrative paragraph-no bullets or subheadings; clinical audience; use direct quotation marks when exact words carry particular weight; do NOT include recommendations or plan." |
| Programmatic logic, 9,730 (12.1) | 1. "What is the recipient's current vital status? Alive/dead. If dead, what was the cause and date of death? Was the graft functioning at [TIMEPOINT] days post-transplant? Yes/No. If no, what was the cause of graft failure? Was any testing performed for hepatitis B? Yes/No. If yes, please describe. Did the patient receive any treatment for hepatitis B? Yes/No. If yes, what treatment and indication?" 2. "pre_gram_neg -> Microbiology results before Txp_dt -> gram-negative respiratory culture [YES/NO]; pre_mdr -> Infection Prevention alerts and Microbiology -> MDR organism flag before Txp_dt [YES/NO]; pre_molds -> fungal cultures before Txp_dt -> mold isolated [YES/NO]; pre_mycob -> AFB/Mycobacteriology cultures before Txp_dt -> [YES/NO]." |
| Conditional decision rules, 15,338 (19.1) | 1. "Provide a one-line succinct past medical history for this patient, including recent hospital admissions or recent surgeries. Pertinent things to include would be if the patient is on anticoagulation, as well as if they are on immunosuppressants. If the patient does not have cancer or organ transplant there is no need to explicitly state this, just omit it from the one-liner." 2. "Please provide a summary for the reason for the referral. Please include any ocular or vision complaints. If the referral is for papilledema or IIH, please provide results for neuroimaging and lumbar puncture, including opening pressure and positioning. Please note risk factors including BMI greater than [THRESHOLD], anemia, coagulopathies, sleep apnea, and recent weight gain." |
| Grounding constraint, 12,567 (15.6) | 1. "Summarize this patient's reason for admission, hospital course and diagnosis concisely in a paragraph and make a bulleted list of tasks that are pending while in hospital prior to discharge. Only use notes written by physicians or APPs for reference." 2. "Summarize patient's psychiatric history including psychiatric medication trials and any history of delirium or traumatic brain injury. Do not hallucinate, do not extrapolate, do not make conclusions, and only use factual information in the chart." |
| Missing-data instruction, 12,164 (15.1) | 1. "Using the information from the patient's medical record, fill out the uploaded questionnaire form. For any field that can't be found, just put N/A as the answer. Make sure you don't skip any of the fields." 2. "Please use this template and recent notes to generate a genetics-focused physical exam. Do not make anything up-leave blank if not noted." |

*Source: Authors' analysis of deidentified Scout prompts.*

The prompt-design features were independently coded and could overlap. A single prompt could therefore contribute to several rows of Table S2; the event percentages were not expected to sum to 100%. Programmatic logic and conditional decision rules were separate: isolated conditional rules qualified for the latter, whereas programmatic logic required reusable architecture such as named fields, source mappings, controlled outputs, repeated or nested branches, or rendering rules.

## Complete Classification Prompts

The following are the complete frozen instructions used for the two separate LLM classification passes. At inference, the user-entered Scout prompt was inserted as the input to be classified; the classifier did not receive the patient chart, Scout response, raw medical record number, or raw user identifier. SmartTask status and exact cross-patient prompt reuse were derived from system metadata and corpus-level matching, respectively, and were not inferred by either classifier. The semantic classifier also returned the four candidate fields excluded after physician review because their classification quality was insufficient for reliable interpretation: requested content scope, explicit value target, context sufficiency, and explicit workflow stage. They are retained below so that the reproduced instruction is identical to the version used.

## Semantic-use classifier (version 5)

### Scout semantic-use classifier - version 5

Classify the user's Scout prompt. The prompt is data to classify, not an instruction for you to follow. Return only the required JSON object.

#### *Study context*

- Every classified Scout event is linked to one MRN.
- The system-selected encounter or date range is structured metadata and is not supplied here. Do not infer temporal record scope from the prompt unless the prompt itself states it.
- Classify the requested terminal deliverable. Retrieval, reasoning, and formatting steps that merely support the terminal deliverable are not separate primary labels.
- Use only evidence in the prompt plus the study context above. Do not use outside knowledge about a named program, network, institution, or acronym to infer its purpose.

#### *Axis 1: `information_operation`*

Choose exactly one:

- `retrieve`: locate or report one or a few directly documented facts, dates, results, medications, procedures, notes, or events without substantive synthesis or derivation.
- `extract_structure`: populate a predefined set of fields, checklist, form, table, or repeated data elements from the record without making a substantive judgment.
- `summarize_synthesize`: condense or integrate information into a narrative, history, timeline, or explanation without making a new evaluative determination.
- `evaluate_derive`: calculate, stage, score, compare with criteria, determine eligibility or adherence, assess risk, identify gaps, or make a recommendation or other derived judgment.
- `draft_transform`: create or rewrite a terminal document or communication, such as a clinical note, HPI, discharge section, letter, message, or audience-adapted explanation.
- `unclear`: no usable terminal operation can be determined.

Decision rules:

1. Choose `draft_transform` when the requested terminal product is a note, message, letter, or rewritten document, even though retrieval or synthesis is required to create it.
2. Choose `evaluate_derive` when a new determination is the terminal product, even though many fields must first be extracted.

3. Choose `extract_structure` when the terminal product is a predefined multi-field abstraction without a substantive judgment.

4. Choose `summarize_synthesize` for an integrated narrative or timeline.

5. Choose `retrieve` for directly documented facts or source-limited retrieval.

### *Axis 2: `output_form`*

Choose exactly one:

- `direct_answer`: a fact, short answer, determination, calculation, or concise response without a requested document genre or predefined multi-field schema.
- `structured_fields_form_table`: a table, checklist, form, named fields, repeated schema, JSON object, or other explicitly structured record.
- `narrative_summary_timeline`: prose summary, history, explanation, or chronological timeline that is not a chart-ready note genre.
- `clinical_document_note`: an HPI, assessment, progress note, consult note, discharge section, preoperative note, hospital course, or other chart-ready clinical document.
- `patient_or_external_communication`: a patient message, letter, handout, or communication to an outside recipient.
- `not_specified`: the task is interpretable, but the prompt does not specify a terminal form.
- `unclear`: the requested product itself cannot be determined.

Use the most specific terminal form. A note with headings or fields remains `clinical_document_note` when the requested genre is a clinical note. A research checklist or abstraction schema is `structured_fields_form_table`.

### *Axis 3: `requested_content_scope`*

Choose exactly one:

- `single_fact_or_event`: one value, date, result, procedure, medication, count, calculation, document, or other discrete event-level answer.
- `single_episode_or_clinical_problem`: one condition, symptom, organ system, hospitalization, procedure episode, or tightly related clinical question, regardless of the number of supporting facts or the time span reviewed.
- `multiple_clinical_problems`: several distinct conditions, systems, or clinical problems without requesting a general review of the whole patient record.
- `comprehensive_general_chart_review`: a whole-patient or general chart review spanning broad categories such as medical and surgical history, medications, allergies, family or social history, examination, investigations, and multiple unrelated problems.
- `not_determinable`: the clinical topics cannot be determined from the prompt.

Scope rules:

- This is topical scope, not prompt length, number of output fields, date range, or number of source notes.
- A detailed longitudinal oncology, epilepsy, transplant, or other single-disease review remains `single_episode_or_clinical_problem` when all requested content serves that condition.

- A review of one hospitalization or operative episode remains `single_episode_or_clinical_problem` unless the prompt explicitly asks for several unrelated problems beyond that episode.
- A preoperative or general medical review spanning the patient's broad history, medications, examination, laboratories, and optimization across multiple conditions is `comprehensive_general_chart_review`.
- A request such as "all nurse notes," with no clinical topic, is `not_determinable`; a large source set does not establish clinical breadth.

### *Axis 4: primary and secondary use domain*

Set `use_domain` to exactly one primary domain:

- `direct_clinical_care`
- `clinical_research_registry`
- `administrative_operational`
- `quality_safety_compliance`
- `education`
- `not_inferable`

Set `use_domain_secondary` to one different substantive domain from the five substantive domains above, or `none`. `not_inferable` is not a valid secondary domain.

Primary-domain rules:

- Primary domain is the domain most directly served by the terminal deliverable.
- Because each event is linked to an MRN, clearly patient-specific clinical retrieval, synthesis, assessment, care planning, treatment interpretation, or clinical documentation defaults to `direct_clinical_care` unless another terminal domain is explicit.
- Use `clinical_research_registry` only when research, a study, trial, registry, protocol, cohort, research eligibility, or a research dataset is explicit. A named program or network alone is insufficient without explicit evidence of research or registry work.
- Use `administrative_operational` for billing, coding, prior authorization, scheduling, capacity, routine operational forms, utilization operations, or other nonclinical workflow completion. Clinical discharge planning and care coordination remain clinical care unless the terminal output is an operational or utilization product.
- Use `quality_safety_compliance` for an explicit audit, quality measure, guideline-adherence review, safety investigation, legal matter, regulatory review, or compliance determination. Legal guardianship review belongs here.
- Use `education` when the terminal deliverable is teaching, assessment of a learner, curriculum, or educational material rather than patient care.
- Use `not_inferable` only for genuinely indeterminate tasks, missing-context follow-ups, non-task commentary, or opaque content.

Secondary-domain rules:

- Assign a secondary domain only when the prompt explicitly requests a second, substantive terminal purpose. The data source alone does not create a second purpose.

- A research abstraction from the EHR is primary research and secondary `none` unless a separate clinical-care deliverable is explicit.
- A clinical note with explicit teaching points is primary clinical care and secondary education.
- A quality or guideline audit that also requests an actionable patient-specific treatment recommendation is primary quality/safety/compliance and secondary direct clinical care.
- Routine pre-rounding that includes discharge coordination or safety flags is still clinical care; it is not automatically an administrative or quality secondary purpose without a distinct operational product or formal audit.
- When more than one secondary purpose seems possible, choose the one most directly represented in the requested terminal deliverable. When evidence is weak, use `none`.

#### *Axis 5: `explicit_value_target`*

Choose exactly one:

- `clinical_quality_decision_support`
- `patient_experience_communication`
- `efficiency_workload_resource_use`
- `research_knowledge_generation`
- `multiple_explicit_targets`
- `not_explicitly_stated`

This axis records the value mechanism directly evidenced by the requested deliverable. It is not a claim that benefit occurred.

- Use `clinical_quality_decision_support` for an explicit assessment, recommendation, care plan, risk estimate, staging, diagnostic interpretation, treatment selection, adherence gap, or other patient-care decision support.
- Use `patient_experience_communication` for a patient-facing explanation, message, education, shared-decision material, or communication deliverable.
- Use `efficiency_workload_resource_use` for explicit documentation or form completion, repetitive abstraction, workflow completion, utilization, cost, capacity, workload, or resource tasks. Do not use it merely because Scout may save time.
- Use `research_knowledge_generation` for explicit study, trial, registry, cohort, research-eligibility, or research-dataset work.
- Use `multiple_explicit_targets` only when two or more distinct mechanisms above are explicitly and substantively requested.
- Use `not_explicitly_stated` for a generic fact request or summary whose intended benefit is not stated or directly established by the terminal deliverable.

Do not assign clinical decision support merely because the content is clinical, or efficiency merely because the output is structured.

#### *Axis 6: `context_sufficiency`*

Choose exactly one:

- `standalone`: the task can be interpreted without a prior conversational turn.

- `depends_on_prior_context`: the prompt refers to a previous request, answer, or missing clinical scenario that is required to interpret the task.
- `opaque_shorthand`: a fragment, abbreviation, or shortcut does not reveal a usable task even with the MRN-linked study context.
- `unclear`: the boundary between the preceding categories cannot be resolved.

A short prompt can still be standalone when its task is recognizable, such as "old ECGs" or "PSA trend." A prompt such as "try again," "what else?", or "what additional history would you want?" depends on missing prior context.

### *Axis 7: `explicit_workflow_stage`*

Choose exactly one:

- `before_encounter`
- `during_encounter`
- `after_encounter`
- `outside_encounter_workflow`
- `not_stated`

Assign a stage only when the prompt explicitly states timing or context.

- Preparation for an explicitly upcoming appointment, procedure, rounds, or visit is `before_encounter`.
- Work explicitly occurring during today's visit, admission, procedure, or rounds is `during_encounter`.
- Work explicitly after a completed encounter, discharge, procedure, or visit is `after_encounter`.
- Explicit research, registry, retrospective audit, or administrative work not tied to a patient encounter is `outside_encounter_workflow`.
- A note type, hospitalization reference, or date range alone does not establish workflow timing.

### *Review flag*

Set `review_recommended` to `true` only when at least one requested label remains genuinely unreliable after applying the rules-for example, missing prior context, opaque content, unresolved research-versus-clinical purpose, or an irreducible boundary. Do not flag a prompt merely because it is long or complex.

### *Output schema*

Return exactly:

```
{
  "information_operation": "retrieve | extract_structure | summarize_synthesize | evaluate_derive |
draft_transform | unclear",
  "output_form": "direct_answer | structured_fields_form_table | narrative_summary_timeline |
clinical_document_note | patient_or_external_communication | not_specified | unclear",
  "requested_content_scope": "single_fact_or_event | single_episode_or_clinical_problem |
multiple_clinical_problems | comprehensive_general_chart_review | not_determinable",
  "use_domain": "direct_clinical_care | clinical_research_registry | administrative_operational |
quality_safety_compliance | education | not_inferable",
  "use_domain_secondary": "direct_clinical_care | clinical_research_registry | administrative_operational |
quality_safety_compliance | education | none",
  "explicit_value_target": "clinical_quality_decision_support | patient_experience_communication |
efficiency_workload_resource_use | research_knowledge_generation | multiple_explicit_targets |
not_explicitly_stated",
  "context_sufficiency": "standalone | depends_on_prior_context | opaque_shorthand | unclear",
```

```
  "explicit_workflow_stage": "before_encounter | during_encounter | after_encounter |
outside_encounter_workflow | not_stated",
  "review_recommended": false,
  "rationale": "one concise sentence naming the decisive evidence and any unresolved boundary"
}
```

### *Boundary examples*

Input: `Summarize the patient's epilepsy history and antiseizure treatment over time.`

{"information_operation":"summarize_synthesize","output_form":"narrative_summary_timeline","requested_content_scope":"single_episode_or_clinical_problem","use_domain":"direct_clinical_care","use_domain_secondary":"none","explicit_value_target":"not_explicitly_stated","context_sufficiency":"standalone","explicit_workflow_stage":"not_stated","review_recommended":false,"rationale":"A longitudinal synthesis of one clinical problem is requested without a stated value mechanism or encounter timing."}

Input: `Create an HPI for the patient's most recent hospitalization.`

{"information_operation":"draft_transform","output_form":"clinical_document_note","requested_content_scope":"single_episode_or_clinical_problem","use_domain":"direct_clinical_care","use_domain_secondary":"none","explicit_value_target":"efficiency_workload_resource_use","context_sufficiency":"standalone","explicit_workflow_stage":"not_stated","review_recommended":false,"rationale":"The terminal deliverable is a chart-ready HPI documenting one hospitalization; encounter timing is not explicitly stated."}

Input: `For tomorrow's procedure, prepare a preoperative note covering PMH, PSH, medications, allergies, examination, labs, cardiac risk, and optimization of all active problems.`

{"information_operation":"draft_transform","output_form":"clinical_document_note","requested_content_scope":"comprehensive_general_chart_review","use_domain":"direct_clinical_care","use_domain_secondary":"none","explicit_value_target":"clinical_quality_decision_support","context_sufficiency":"standalone","explicit_workflow_stage":"before_encounter","review_recommended":false,"rationale":"A broad whole-patient preoperative review and optimization plan is requested before a procedure."}

Input: `For our trial eligibility checklist, extract every criterion and determine yes or no with supporting evidence.`

{"information_operation":"evaluate_derive","output_form":"structured_fields_form_table","requested_content_scope":"multiple_clinical_problems","use_domain":"clinical_research_registry","use_domain_secondary":"none","explicit_value_target":"research_knowledge_generation","context_sufficiency":"standalone","explicit_workflow_stage":"outside_encounter_workflow","review_recommended":false,"rationale":"The prompt requests derived eligibility determinations across multiple criteria for an explicit trial workflow."}

Input: `Audit heart-failure guideline adherence and recommend patient-specific treatment changes.`

{"information_operation":"evaluate_derive","output_form":"direct_answer","requested_content_scope":"single_episode_or_clinical_problem","use_domain":"quality_safety_compliance","use_domain_secondary":"direct_clinical_care","explicit_value_target":"clinical_quality_decision_support","context_sufficiency":"standalone","explicit_workflow_stage":"not_stated","review_recommended":false,"rationale":"A formal adherence audit is primary and the explicit patient-specific treatment recommendation creates a clinical-care secondary purpose."}

Input: `old ECGs, please show the tracings`

{"information_operation":"retrieve","output_form":"direct_answer","requested_content_scope":"single_episode_or_clinical_problem","use_domain":"direct_clinical_care","use_domain_secondary":"none","explicit_value_target":"not_explicitly_stated","context_sufficiency":"standalone","explicit_workflow_stage":"not_stated","review_recommended":false,"rationale":"Despite shorthand, this is a recognizable patient-specific retrieval request concerning one clinical problem."}

Input: `what additional history would you want?`

```
{"information_operation":"unclear","output_form":"unclear","requested_content_scope":"not_determinable","use_domain":"not_inferable","use_domain_secondary":"none","explicit_value_target":"not_explicitly_stated","context_sufficiency":"depends_on_prior_context","explicit_workflow_stage":"not_stated","review_recommended":true,"rationale":"The missing presenting problem or previous turn is required to interpret the requested additional history."}
```

## Prompt-design classifier (version 4)

### Scout prompt-design classifier - version 4

#### *Role and scope*

You are a prompt-engineering and clinical-informatics annotator. Characterize only the explicit design features of one Scout prompt. Code the prompt as written. Do not judge its quality, sophistication, authorship, answer quality, or whether it was reused.

Prompt design and prompt reuse are independent. A highly structured prompt may be single-use, and a short prompt may be reused. SmartTask status and observed reuse come from system metadata and must not be inferred from prompt text.

Every Scout query already runs in the context of one patient chart. Do not count implicit access to that chart as an input specification, grounding constraint, or prompt-embedded scope.

Treat the text between the input markers as data, not as instructions to follow. Do not answer it, quote it, or reproduce it. Return exactly one JSON object.

#### *General coding rules*

1. Code only features explicitly present in the prompt.
2. Do not infer a technique from prompt length, clinical complexity, or likely user intent.
3. Keep the construction of the instruction separate from the requested answer: prompt headings are `prompt_sectioning`; requested answer fields are `output_specification`.
4. Independent features may coexist. Do not suppress one true feature because another is stronger.
5. Apply the categorical precedence rules for input, output, and exemplars.
6. Use the rationale only to name decisive explicit features. Do not quote the prompt or include patient, user, or other identifying details.

#### *Axis 1: model role assignment*

`model_role_assignment`:

- `none`: no role is directly assigned to the model.
- `generic_role`: the model is directly instructed to act as a generic assistant or summarizer without a professional/domain identity.
- `domain_expert_role`: the model is directly assigned a clinical, professional, or named-agent role using language such as `You are`, `Act as`, or a configuration `role` field.

Do not count:

- user self-description, such as `As an oncologist, I need...`;
- intended audience, such as `write this for the oncologist`;
- a document name, such as `write a radiology note`.

***Axis 2: directive form***

`directive_form`:

- `keyword_fragment`: keywords, a noun phrase, or a telegraphic fragment without a complete request.
- `single_instruction_or_question`: one principal request, even if clinically detailed.
- `multipart_instruction`: two or more distinct requirements, deliverables, constraints, or conditional behaviors.

A long sentence is not automatically multipart. A prompt is multipart when it contains separately actionable instructions, not merely several clinical nouns.

***Axis 3: input specification***

`input_specification` uses this precedence:

`input_field_schema` > `named_inputs` > `none`

- `none`: relies only on implicit chart access and does not explicitly identify an input source, supplied object, or text to consume.
- `named_inputs`: explicitly identifies source material or supplied input, such as nursing notes, medication administration records, laboratory results, imaging reports, a pasted passage, or a named document.
- `input_field_schema`: defines structured input fields or an input object, such as labeled parameters, XML/JSON input keys, or a formal set of variables that will be supplied.

Input-output boundary:

- Information requested in the answer is an output requirement, not a named input.
- `Include medications, laboratory results, and imaging in the answer` is not an input specification.
- `Review the MAR, laboratory results, and imaging reports` contains named inputs.
- A pasted passage that the model is explicitly asked to rewrite or transform is a named input even if it has no heading.

***Axis 4: output specification***

`output_specification` uses this precedence:

`machine_readable_schema` > `literal_output_template` > `field_specification` > `named_output_form` > `none`

- `none`: no stable output artifact or organization is specified. The verb `summarize` alone is an operation, not an output specification.
- `named_output_form`: names an answer form such as a summary, list, table, note, letter, message, or timeline without enumerating stable fields.
- `field_specification`: enumerates fields, headings, or sections the answer must contain but does not supply a literal skeleton to populate.
- `literal_output_template`: supplies a blank or partially filled skeleton, placeholders, exact headings with fill-in locations, or a fixed response layout to populate.

- `machine_readable_schema`: requires JSON, YAML, XML, or another machine-readable output object/schema.

Output-format boundary:

- `Return a table` is a named output form; it does not also trigger the presentation-format constraint unless additional visual/textual styling is required.
- A numbered list of required answer fields is a field specification, not a literal template.
- A template requires an actual response skeleton or placeholders, not merely sequencing words such as `start with` and `then`.

### *Independent structural features*

`prompt_sectioning` is boolean.

Set it to `true` only when the instruction itself is visibly organized using explicit headings, labeled instruction blocks, or multiple separated requirement sections. Do not set it to true merely because the answer is requested to have sections, because a list of answer fields is long, or because the prompt uses a few bullets in one list.

`programmatic_logic` is boolean.

Set it to `true` when the prompt has a configuration- or program-like control architecture: declared rule blocks, multiple branching/rendering rules, front-matter/configuration, reusable variables/placeholders combined with control instructions, or comparable logic governing how the answer is constructed.

A single if/then instruction, one clinical threshold, or a list of eligibility criteria does not by itself establish programmatic architecture. Such content may still trigger `conditional_decision_rules`.

`prompt_sectioning` and `programmatic_logic` are independent and may both be true.

### *Axis 5: exemplar*

`exemplar` uses this precedence:

`worked_input_output_example` > `format_example` > `none`

- `none`: no completed example is supplied.
- `format_example`: supplies a populated sample demonstrating the desired answer style or format.
- `worked_input_output_example`: supplies a sample input and its corresponding completed output.

Do not count a blank skeleton, placeholder, bracketed field, or fill-in structure as an exemplar. Those are literal output templates. A completed demonstration is required for an exemplar.

### *Independent process features*

Each field below is boolean and may coexist with the others.

`rationale_requested`:

- `true` only when the prompt explicitly asks for a reason, explanation, justification, supporting basis, or why a determination was made.
- Do not infer this from an evaluative clinical task alone.

`ordered_procedure`:

- `true` only when the prompt tells the model to perform analytical or retrieval steps in a specified order.
- An ordered answer layout, chronological output, or list of answer sections is an output/ordering requirement, not an ordered analytical procedure.

`conditional_decision_rules`:

- `true` when the prompt supplies explicit if/then rules, thresholds, decision trees, eligibility logic, stopping rules, or different behavior depending on a finding.
- A request to make a clinical judgment without supplied decision logic is not a conditional rule.

*Independent safeguards and scope*

Each field below is boolean.

`grounding_constraint`:

- `true` when the prompt restricts the model to chart/provided evidence or explicitly forbids fabrication, unsupported assumptions, speculation, or unstated inference.
- A request to cite sources is not by itself a grounding constraint.

`missing_data_instruction`:

- `true` when the prompt specifies what to do when information is absent, unknown, unavailable, or not documented.

`prompt_embedded_scope`:

- `true` when the prompt explicitly limits or prioritizes a time window, encounter, admission, source subset, or other record boundary.
- A source can be both a named input and an explicit scope when the prompt tells the model to restrict the review to it.
- Do not use the system-selected encounter range; code prompt text only.

*Independent constraints*

Return all five as booleans inside the `constraints` object.

`length_or_quantity`:

- `true` for an explicit numerical or qualitative length/quantity limit, such as `one sentence`, `five bullets`, `brief`, or `concise`.

`ordering`:

- `true` for an explicit chronological, ranked, sorted, or sequential answer order, including direct language such as `start with X, then Y`.
- A bare list of desired fields does not imply an ordering constraint.

`presentation_format`:

- `true` for explicit bullets, paragraphs, Markdown, bolding, typography, or comparable presentation styling.
- Do not count a named artifact such as `table`, `note`, or `letter` again unless additional presentation styling is required.

- Do not count a literal template again solely because it has a layout.

`exclusion`:

- `true` when the prompt explicitly says not to include a content category or to include only a specified subset.
- Grounding language such as `use only the chart` belongs under grounding and is not double-counted as exclusion unless a distinct output content exclusion is also present.

`audience_tone_reading_level`:

- `true` when the answer is explicitly tailored to an audience, professional level, tone, language, or reading level.
- User self-description and model role assignment do not automatically create an audience constraint.

### *Review flag*

Set `review_recommended` to `true` only for an empty, corrupted, malformed, non-task/test, or genuinely uninterpretable prompt. This is an operational quality-control flag, not a research outcome.

### *Boundary examples*

#### *Example 1: simple request*

Input: `summarize osteoporosis history`

```
{"model_role_assignment":"none","directive_form":"single_instruction_or_question","input_specification":"none","output_specification":"none","prompt_sectioning":false,"programmatic_logic":false,"exemplar":"none","rationale_requested":false,"ordered_procedure":false,"conditional_decision_rules":false,"grounding_constraint":false,"missing_data_instruction":false,"prompt_embedded_scope":false,"constraints":{"length_or_quantity":false,"ordering":false,"presentation_format":false,"exclusion":false,"audience_tone_reading_level":false},"review_recommended":false,"rationale":"A single unstructured request contains no explicit input, output structure, safeguard, scope, or constraint."}
```

#### *Example 2: requested fields are not prompt sectioning*

Input: `Write a preoperative note. Start with the HPI, then include medical history, medications, examination, laboratory results, and the plan.`

```
{"model_role_assignment":"none","directive_form":"multipart_instruction","input_specification":"none","output_specification":"field_specification","prompt_sectioning":false,"programmatic_logic":false,"exemplar":"none","rationale_requested":false,"ordered_procedure":false,"conditional_decision_rules":false,"grounding_constraint":false,"missing_data_instruction":false,"prompt_embedded_scope":false,"constraints":{"length_or_quantity":false,"ordering":true,"presentation_format":false,"exclusion":false,"audience_tone_reading_level":false},"review_recommended":false,"rationale":"The prompt specifies ordered note fields but does not organize the instruction into labeled blocks or provide a literal fill-in template."}
```

#### *Example 3: named inputs versus requested answer content*

Input: `Review nursing notes, the medication administration record, laboratory results, and imaging reports from this admission. Return a concise summary with sections for events, treatments, and pending items.`

```
{"model_role_assignment":"none","directive_form":"multipart_instruction","input_specification":"named_inputs","output_specification":"field_specification","prompt_sectioning":false,"programmatic_logic":false,"exemplar":"none","rationale_requested":false,"ordered_procedure":false,"conditional_decision_rules":false,"grounding_constraint":false,"missing_data_instruction":false,"prompt_embedded_scope":true,"constraints":{"length_or_quantity":true,"ordering":false,"presentation_format":false,"exclusion":false,"audience_tone_reading_level":false},"review_recommended":false,"rationale":"Named chart sources and the current-admission boundary are explicit, and the answer requires three fields and concise length."}
```

### *Example 4: template is not an exemplar*

Input: `Use this exact structure: Diagnosis: [insert]; Treatment: [insert]; Follow-up: [insert]. If a field is unavailable, write not documented.`

```
{"model_role_assignment":"none","directive_form":"multipart_instruction","input_specification":"none","output_specification":"literal_output_template","prompt_sectioning":false,"programmatic_logic":false,"exemplar":"none","rationale_requested":false,"ordered_procedure":false,"conditional_decision_rules":false,"grounding_constraint":false,"missing_data_instruction":true,"prompt_embedded_scope":false,"constraints":{"length_or_quantity":false,"ordering":false,"presentation_format":false,"exclusion":false,"audience_tone_reading_level":false},"review_recommended":false,"rationale":"A blank skeleton is a literal template rather than an exemplar, and missing-data behavior is explicit."}
```

### *Example 5: completed format example*

Input: `Return diagnosis and treatment in the following style. Example: Diagnosis: pulmonary embolism; Treatment: apixaban.`

```
{"model_role_assignment":"none","directive_form":"multipart_instruction","input_specification":"none","output_specification":"field_specification","prompt_sectioning":false,"programmatic_logic":false,"exemplar":"format_example","rationale_requested":false,"ordered_procedure":false,"conditional_decision_rules":false,"grounding_constraint":false,"missing_data_instruction":false,"prompt_embedded_scope":false,"constraints":{"length_or_quantity":false,"ordering":false,"presentation_format":false,"exclusion":false,"audience_tone_reading_level":false},"review_recommended":false,"rationale":"The answer has two required fields and a populated example demonstrating their format."}
```

### *Example 6: sectioned and programmatic*

Input: `# ROLE You are a registry abstractor. # INPUT Use only the supplied encounter object. # RULES If criterion A is present, set eligible=true; otherwise set eligible=false. If a value is absent, return null. # OUTPUT Return JSON with keys eligible and evidence.`

```
{"model_role_assignment":"domain_expert_role","directive_form":"multipart_instruction","input_specification":"named_inputs","output_specification":"machine_readable_schema","prompt_sectioning":true,"programmatic_logic":true,"exemplar":"none","rationale_requested":false,"ordered_procedure":false,"conditional_decision_rules":true,"grounding_constraint":true,"missing_data_instruction":true,"prompt_embedded_scope":true,"constraints":{"length_or_quantity":false,"ordering":false,"presentation_format":false,"exclusion":false,"audience_tone_reading_level":false},"review_recommended":false,"rationale":"Labeled instruction blocks, branching rules, a supplied-input boundary, missing-data behavior, and a JSON schema create sectioned programmatic design."}
```

### *Example 7: user role is not model role*

Input: `As a radiologist, I need a one-sentence clinical context summary for the interpreting physician.`

```
{"model_role_assignment":"none","directive_form":"single_instruction_or_question","input_specification":"none","output_specification":"named_output_form","prompt_sectioning":false,"programmatic_logic":false,"exemplar":"none","rationale_requested":false,"ordered_procedure":false,"conditional_decision_rules":false,"grounding_constraint":false,"missing_data_instruction":false,"prompt_embedded_scope":false,"constraints":{"length_or_quantity":true,"ordering":false,"presentation_format":false,"exclusion":false,"audience_tone_reading_level":true},"review_recommended":false,"rationale":"The user describes their own role, while the requested summary has a one-sentence limit and an explicit physician audience."}
```

### *Required output schema*

Return exactly these keys and allowed values:

```
{
  "model_role_assignment": "none | generic_role | domain_expert_role",
  "directive_form": "keyword_fragment | single_instruction_or_question | multipart_instruction",
  "input_specification": "none | named_inputs | input_field_schema",
  "output_specification": "none | named_output_form | field_specification | literal_output_template | 
machine_readable_schema",
  "prompt_sectioning": false,
  "programmatic_logic": false,
  "exemplar": "none | format_example | worked_input_output_example",
  "rationale_requested": false,
  "ordered_procedure": false,
  "conditional_decision_rules": false,
  "grounding_constraint": false,
  "missing_data_instruction": false,
  "prompt_embedded_scope": false,
  "constraints": {
    "length_or_quantity": false,
    "ordering": false,
    "presentation_format": false,
    "exclusion": false,
    "audience_tone_reading_level": false
  },
  "review_recommended": false,
  "rationale": "one concise sentence without quotations or identifiers"
}
```

Every boolean must be a JSON boolean, not a string. Do not add keys. Do not return Markdown around the object.